\documentclass[twocolumn]{aastex631}

\usepackage{amsmath}
\usepackage{mathtools}
\usepackage{url}

\shorttitle{Dark Comets?}
\shortauthors{Seligman et al.}

\begin{document}

\title{Dark Comets? Unexpectedly Large Nongravitational Accelerations on a Sample of Small Asteroids }

\author[0000-0002-0726-6480]{Darryl Z. Seligman}
\affiliation{Department of Astronomy and Carl Sagan Institute, Cornell University, 122 Sciences Drive, Ithaca, NY, 14853, USA}

\correspondingauthor{Darryl Z. Seligman}
\email{dzs9@cornell.edu}

\author[0000-0003-0774-884X]{Davide Farnocchia}
\affiliation{Jet Propulsion Laboratory, California Institute of Technology,
4800 Oak Grove Dr., Pasadena, CA 91109, USA}

\author[0000-0001-7895-8209]{Marco Micheli}
\affiliation{ESA NEO Coordination Centre, Largo Galileo Galilei 1,
 I-00044 Frascati (RM), Italy}
 
 \author[0000-0002-6034-5452]{David Vokrouhlick\'y}
\affiliation{Institute of Astronomy, Charles University, 
 V Hole\v{s}ovi\v{c}k\'ach 2, CZ-18000 Prague 8, Czech Republic}
 
 \author[0000-0002-0140-4475]{Aster G. Taylor}
\affiliation{Dept. of Astronomy and Astrophysics, University of Chicago, 5640 S Ellis Ave, Chicago, IL 60637, USA} 

 \author[0000-0003-3240-6497]{Steven R. Chesley}
\affiliation{Jet Propulsion Laboratory, California Institute of Technology, 
4800 Oak Grove Dr., Pasadena, CA 91109, USA}
 \author[0000-0002-8716-0482]{Jennifer B. Bergner}
\affiliation{University of Chicago, Department of the Geophysical Sciences, Chicago, IL 60637, USA}

\author[0000-0002-5396-946X]{Peter Vere\v{s}}
\affiliation{Harvard-Smithsonian Center for Astrophysics, Minor Planet Center, 60 Garden Street, Cambridge, MA 02138, USA}

\author[0000-0001-6952-9349]{Olivier R. Hainaut}
\affiliation{European Southern Observatory, Karl-Schwarzschild-Strasse 2, Garching bei München, D-85748, Germany}

\author[0000-0002-2058-5670]{Karen J. Meech}
\affiliation{Institute for Astronomy, University of Hawaii, 
2680 Woodlawn Dr., Honolulu, HI 96822, USA}

\author[0000-0002-6509-6360]{Maxime Devogele}
\affiliation{Arecibo Observatory, University of Central Florida, HC-3 Box 53995, Arecibo, PR 00612, USA}

\author{Petr Pravec}
\affiliation{Astronomical Institute, Academy of Sciences of the Czech Republic, Fri\v{c}ova 1, CZ-25165 Ond\v{r}ejov, Czech Republic}

\author{Rob Matson}

\author{Sam Deen}
\affiliation{Independent Researcher}

\author{David J. Tholen}
\affiliation{Institute for Astronomy, University of Hawaii, 2680 Woodlawn Drive, Honolulu, HI 96822, USA}

\author[0000-0002-0439-9341]{Robert Weryk}
\affiliation{Physics and Astronomy, The University of Western Ontario, 1151 Richmond Street, London ON N6A 3K7, Canada}

\author[0000-0002-4042-003X]{Edgard G. Rivera-Valent\'{i}n}
\affiliation{Johns Hopkins University Applied Physics Laboratory, 11100 Johns Hopkins Road, Laurel, MD 20723, USA.}

\author[0000-0003-1383-1578]{Benjamin N. L. Sharkey}
\affiliation{Lunar and Planetary Laboratory, University of Arizona, 1629 E University Blvd, Tucson, AZ 85721, USA}

\begin{abstract}
We report statistically significant detections of non-radial nongravitational accelerations based on astrometric data in the photometrically inactive  objects 1998 KY$_{26}$, 2005 VL$_1$, 2016 NJ$_{33}$, 2010 VL$_{65}$, 2016 RH$_{120}$, and 2010 RF$_{12}$. The magnitudes of the nongravitational accelerations are greater than those typically induced by the Yarkovsky effect and there is no radiation-based, non-radial effect that can be so large. Therefore, we hypothesize that the accelerations are driven by outgassing, and calculate implied H$_2$O production rates for each object. We attempt to reconcile  outgassing induced acceleration with the lack of visible comae or photometric activity via the absence of surface dust and low levels of gas production. Although these objects are small and some are rapidly rotating,  surface cohesive forces  are  stronger than the rotational forces and rapid rotation alone cannot explain the lack of surface debris. It is possible that surface dust was  removed previously, perhaps via outgassing activity that increased the rotation rates to their present day value.  We calculate dust production rates of order $\sim10^{-4}$ g s$^{-1}$ in each object assuming that the nuclei are bare, within the upper limits of dust production from a sample stacked image of 1998 KY$_{26}$ of $\dot{M}_{\rm Dust}<0.2$ g s$^{-1}$. This production corresponds to  brightness variations of order $\sim0.0025\%$, which are undetectable in extant photometric data. We assess the future observability of each of these targets, and find that the orbit of 1998 KY$_{26}$ --- which is also the target for the extended Hayabusa2 mission --- exhibits favorable viewing geometry before 2025.
\end{abstract}


\section{Introduction} \label{sec:intro}

A typical --- albeit simplistic --- classification of Solar System small bodies is to categorize the populations based on volatile-driven activity. In this classical picture, comets are defined as icy objects that produce dusty comae and presumably formed in distant regions of the Solar System, while  asteroids lack volatiles due to prolonged exposure to  solar irradiation. However, recent advances have demonstrated that this simplistic classification may not be an accurate depiction of the  census of small bodies within the Solar System. A subset of objects on cometary orbits lack detectable activity, while some asteroids have displayed volatile-induced comae. These intriguing continuum objects could offer insights into little understood processes such as cometary fading \citep{Brasser2015} and volatile delivery to terrestrial planets \citep{Chyba1990,Owen1995,Albarede2009}.

It is generally thought that asteroids can be grouped into classes based on their sizes.  The typical timescale for a $\sim$10 km scale asteroid to experience a catastrophic impact is approximately the age of the Solar System \citep{Bottke2005,Bottke2015}. It is therefore believed that asteroids  larger than $\sim$10 km are intact primordial remnants, while smaller objects  with diameter between 200 m to $\sim$10 km are  re-accumulated rubble-piles held together by self-gravity   \citep{Harris1979,Harris1979b,Harris1996,Walsh2018}.  As evidence, the $\sim17$ km equivalent diameter asteroid Eros  visited by   NEAR-Shoemaker displayed geologic properties consistent with being a damaged but intact primordial remnant \citep{Cheng2002}. Notable examples of extremely porous rubble-pile asteroids are (25143) Itokawa \citep{Fujiwara2006}, (162173) Ryugu \citep{Watanabe2019}, and (101955) Bennu \citep{Barnouin2019}.

A subset of asteroids are known to exhibit  activity, also called active asteroids \citep{Jewitt2012,Hsieh2017,Jewitt22_asteroid}. Almost all active asteroids known to date have diameters consistent with being rubble-piles, with the exception of the large objects (1) Ceres and (493) Griseldis.  A subset of active asteroids are  the  Main Belt Comets (MBCs) ---  objects that reside in the main asteroid belt between Mars and Jupiter  and display cometary activity driven by the sublimation of volatiles \citep{Hsieh2006}. The first MBC discovered was Comet 133P/(7968) Elst-Pizarro  \citep{Elst1996,Boehnhardt1996,Toth2000,Hsieh2004} and several others  have since been identified: 238P/Read, 259P/Garradd,  288P/(300163) 2006 VW139, 313P/Gibbs, 324P/La Sagra, 358P/PANSTARRS, 107P/(4015) Wilson-Harrington, and  433P/(248370) 2005 QN173 \citep[for images, see Figure 13 in][]{Jewitt22_asteroid}. 

Targeted, homogeneous searches for  MBCs imply occurrence rates for these objects within the total population between $<1/500$ and $\sim 1/300$ \citep{Sonnet2011,Bertini2011,Snodgrass2017,Ferellec2022}. Indirect measurements of activity can come from anisotropic mass-loss-driven  nongravitational accelerations \citep{Whipple1950,Whipple1951}. For example, \citet{Hui2017} reported statistically significant nongravitational accelerations in the active asteroids 313P/Gibbs, 324P/La Sagra, and (3200) Phaethon. 

 Non-sublimation  effects such as impacts \citep{Snodgrass2010} and rotational effects \citep{Jewitt2014} can also cause activity in asteroids.  The $\sim$6 km sized object (3200) Phaethon was first determined to be active when  its  association with the Geminid meteoroid stream was  identified \citep{Gustafson1989,Williams1993}. Subsequent observations of (3200) Phaethon near its perihelion revealed a small tail with micron sized dust production rates of $\sim 3$ kg s$^{-1}$ \citep{Jewitt2010,Jewitt2013,Li2013,Hui2017b}. Because these production rates are not large enough to explain the meteoroid stream, it has been suggested that processes such as repeated thermally induced stresses \citep{Jewitt2010}, sublimation of minerologically bound sodium \citep{Masiero2021}, rotational effects \citep{Ansdell2014,Nakano2020} and geometric effects \citep{Hanus2016b,Taylor2019} also contribute to the activity of (3200) Phaethon.  The Japan Aerospace Exploration Agency  (JAXA) DESTINY+ mission  will visit (3200) Phaethon and is expected to launch  in 2024 \citep{Arai2021}. DESTINY+ will be equipped with a mass spectrometer which would detect elemental compositions of dust released by the object \citep{Kruger2019}. 
 
 (101955) Bennu is another intriguing case of an active asteroid. The OSIRIS-REx spacecraft observed repeated periods of particle ejection from the $\sim500$ m rubble pile \citep{Lauretta2019,Hergenrother2019}. The average mass loss rate of dust was  measured to be only $\dot{M}_{\rm Dust}\sim10^{-4}$ g s$^{-1}$ \citep{Hergenrother2020}, but the source of the activity is unclear   \citep{Bottke2020,Molaro2020,Chesley2020}.  

A somewhat related but distinct family of minor bodies are the \textit{inactive} comets. For example, ``Manx-comets'' are objects with Long Period Comet (LPC) trajectories that exhibit little or no cometary activity. The  Manx-comet  C/2014 S3 (PANSTARRS) displayed  levels of cometary activity that were 5-6 orders of magnitude less than typical comets, and spectral features similar to    S- type asteroids \citep{Meech2016}. The origin of the Manx-comets are unclear, because Jupiter scatters objects  within the H$_2$O snowline primarily into the interstellar medium and not into the Oort cloud \citep{Hahn1999,Shannon2015}. Other inactive comets are  Damocloids \citep[named after the first object (5335) Damocles; ][]{Asher1994,Jewitt2005} and Asteroids on Cometary Orbits (ACOs), defined as inactive objects on Halley-type comet orbits or Jupiter Family Comet (JFC) orbits, respectively. Damocloids and ACOs  are believed to be comets that have little or no  activity due to cometary fading \citep{Wang2014,Brasser2015}, or the depletion of volatiles or mantling \citep{Podolak1985,Prialnik1988}. Observed spectral features and surface colors of Damocloids and ACOs indicate that they are likely extinct or dormant comets \citep{Jewitt2005,Licandro2018}.

The recently discovered population of interstellar objects also appear to exhibit a continuum of apparent activity. The first interstellar object 1I/`Oumuamua was detected on UT 2017 October 19 \citep{Williams17} and exhibited a mixture of cometary and asteroidal properties. On the other hand, the 0.2-0.5 km radius scale \citep{Jewitt20} 2I/Borisov displayed a visible cometary tail \citep{Jewitt2019,Guzik2020, Kim2020,Cremonese2020,Hui2020,Bodewits2020, yang2021}. The ${\sim}120$ m diameter 1I/`Oumuamua displayed an extreme light curve amplitude \citep{Knight2017,Bolin2017,Fraser2017,McNeill2018,Belton2018,Mashchenko2019,Drahus18}, reddened color \citep{masiero2017,Fitzsimmons2017,Bannister2017,Ye2017} and incoming trajectory consistent with the local standard of rest \citep{mamajek2017,Gaidos2017, Feng2018,Fernandes2018,Hallatt2020,Hsieh2021}. Of particular interest and relevance to this paper, `Oumuamua exhibited no evidence for an extended dust coma in deep stacked composite images \citep{Meech2017,Jewitt2017}. Moreover, it exhibited no obvious infrared fluorescence of carbon-based outgassing species (e.g., CO or CO$_2$) based on observations obtained with the \textit{Spitzer Space Telescope} \citep{Trilling2018}.  However, the trajectory of the object was affected by a  significant nongravitational acceleration \citep{Micheli2018}. \citet{Micheli2018} argued that the most likely cause of this acceleration ---  which was primarily in the radial direction opposite the direction of the sun --- was cometary outgassing. The physical properties of the first two interstellar interlopers are readily summarized; see  \citet{Jewitt2022} and \citet{MoroMartin2022} for recent reviews. 

Recently, Farnocchia et al. (submitted 2022) presented observations that revealed that the object 2003 RM exhibited a significant transverse nongravitational acceleration incompatible with the Yarkovsky effect. However, this object, like `Oumuamua, lacked cometary activity clearly visible in photometric images. They found that outgassing can cause the acceleration and still escape photometric detection. In this  paper, we report identification of similar nongravitational accelerations in five other small bodies on asteroid orbits.

 \begin{figure}
\begin{center}
       \includegraphics[scale=0.35,angle=0]{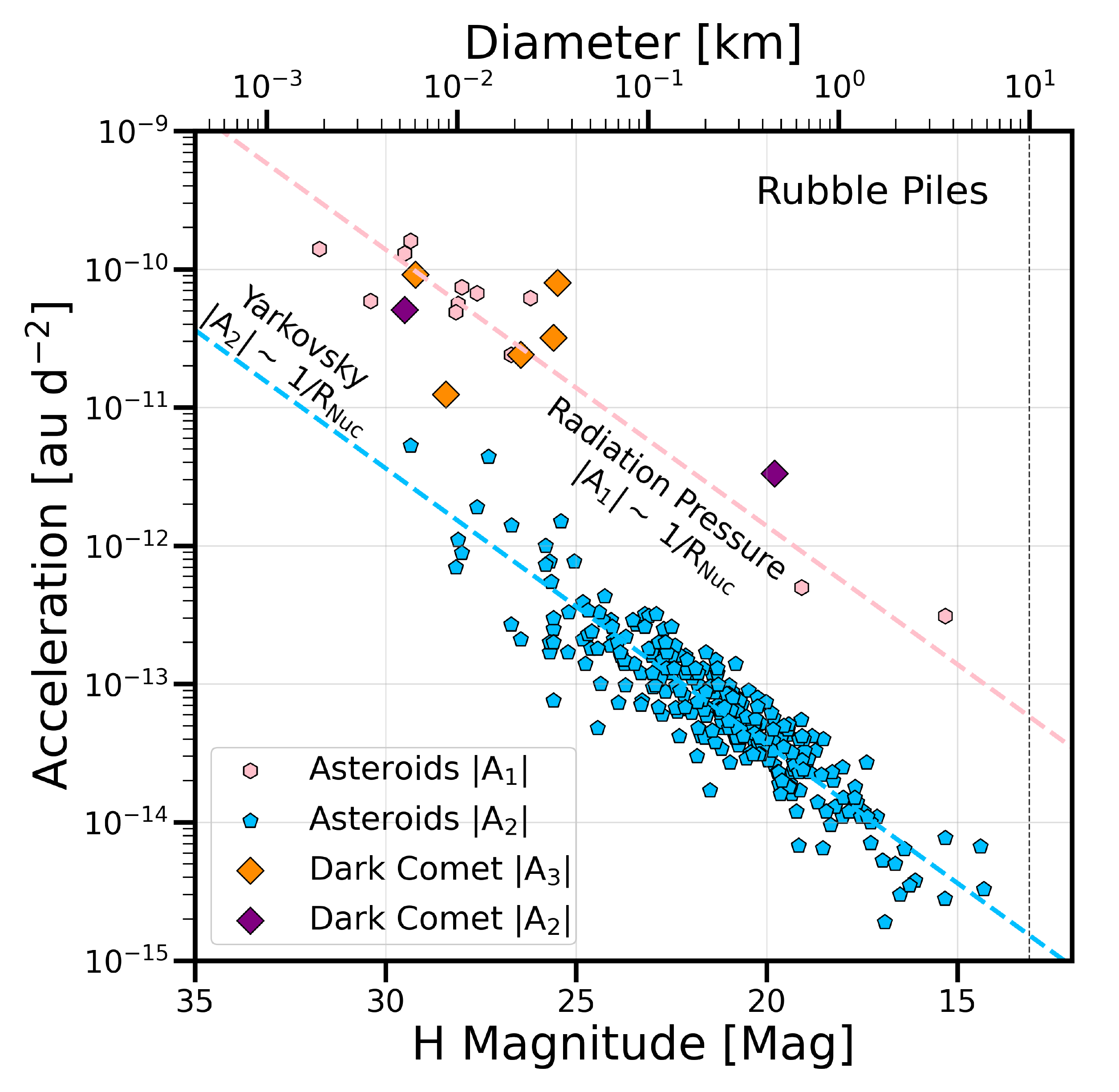}
    \caption{Nongravitational accelerations measured in asteroids versus their absolute magnitude $H$. Measured  radial nongravitational acceleration $A_1$ are due to solar radiation pressure (pink) and  transverse acceleration $A_2$ are due to the Yarkovsky effect (blue).  The out-of-plane nongravitational acceleration $A_3$ measured in the objects presented in this paper are shown in orange diamonds. The blue and pink dashed lines show lines of best fit, assuming that the acceleration is proportional to the inverse of the nuclear size, and computed using the relationship in Equation \ref{eq:diameter_mag} assuming an albedo $p=0.1$. The log-scaled intercepts are -15.858 for the radiation pressure, and -17.440 for the Yarkovsky effect. Dark comet candidates with anomalous $A_2$ values, 2003 RM and 2006 RH$_{120}$, are shown in purple diamonds.} 
 \label{fig:accelerations} 
\end{center}
\end{figure}

\section{Detections of Nongravitational Acceleration}\label{sec:non_grav_detection}

\subsection{Possible Nongravitational Perturbations on Small Bodies}
The trajectories of small bodies are typically estimated from observational datasets,  primarily  optical astrometry and sometimes  radar delay and Doppler measurements.
As the observational arcs extend in time and the accuracy of the data improves, the resulting orbits become better constrained. Therefore,  the requirements on the fidelity of the force model to compute the orbit become more stringent.

It has long been established that  the motion of comets can be significantly perturbed by nongravitational accelerations due to outgassing in addition to  typical gravitational forces. \citet{marsden1973} introduced  a parametric model where the nongravitational acceleration is described as:
\begin{equation}\label{eq:forces}
\mathbf a_{\rm NG} = \bigg( A_1 \hat{\mathbf r} + A_2 \hat{\mathbf t} + A_3 \hat{\mathbf n}\bigg) \, g(r)\,.
\end{equation}
In Equation \ref{eq:forces},  $\hat{\mathbf r}$, $\hat{\mathbf t}$, $\hat{\mathbf n}$ are the orbital radial, transverse, and out-of-plane directions, $g(r)$ is a function based on the H$_2$O sublimation profile capturing the dependence on the heliocentric distance $r$, and $A_1$, $A_2$, and $A_3$ are free parameters that give the acceleration components that the comet would experience at a heliocentric distance of $r=1$ au.

For asteroids, the two main nongravitational perturbations are the Yarkovsky effect \citep{Vokrouhlicky2015_ast4} and solar radiation pressure \citep{Vokrouhlicky2000}.
By setting $g(r) = (1\text{ au}/r)^2$, these two perturbations can be modeled as purely transverse $A_2 g(r)$ and purely radial $A_1 g(r)$ accelerations using the \citet{marsden1973} formalism \citep[e.g.,][]{Farnocchia2015ast4}.
Despite being smaller than those on comets, these two perturbations can cause detectable deviations from a gravity-only trajectory for sufficiently long  observational data arcs.
In particular, the Yarkovsky effect has been detected on more than 200 near-Earth asteroids \citep[e.g.,][]{Greenberg2020} shown in blue points in Figure \ref{fig:accelerations}.
Solar radiation pressure is a radial acceleration and is therefore less effective at producing significant orbital deviations. As such, it has been measured only on
a handful of small asteroids \citep[MPEC 2008-D12,\footnote{\url{https://www.minorplanetcenter.net/mpec/K08/K08D12.html}}][]{Micheli2012,Micheli2013,Micheli2014,Mommert2014bd,Mommert2014md,Farnocchia2017TC25,Fedorets2020} shown in pink points in Figure \ref{fig:accelerations}.
As a testament of the fidelity of the modeling of the force model for asteroids, including these nongravitational forces, \citet{Farnocchia2021} were able to reconstruct the trajectory of asteroid (101955) Bennu by matching ground-based optical and radar astrometry from 1999 to 2018 as well as meter-level ranging measurements from the OSIRIS-REx mission proximity operations from January 2019 to October 2020.

\subsection{Asteroids with Excess Nongravitational Acceleration}
In the accompanying paper Farnocchia et al. (submitted 2022), we report the detection of an anomalously large nongravitational acceleration in the object 2003 RM. In that paper, we conclude that outgassing could have caused the acceleration of 2003 RM while not producing a detectable signature in photometric data. Further, \citet{Chesley2016} reported that the object 2006 RH120, which was temporarely captured by the Earth for about a year starting in June 2006 \citep{Granvik2012}, exhibited a transverse nongravitational acceleration that was inconsistent with typical forces observed acting on asteroids. In this paper, we report  the detection of out-of-plane nongravitational accelerations  in a sample of five other objects that we measured from a combination of serendipitous (and uncoordinated) astrometric observations: 1998 KY$_{26}$, 2005 VL$_1$, 2016 NJ$_{33}$, 2010 VL$_{65}$, and 2010 RF$_{12}$, all identified as point source asteroidal objects with no extended dust coma or evidence for outgassing. We also detect significant out-of-plane perturbations for 2006 RH$_{120}$. There have been no reports of clear detection of cometary activity for any of these small objects to the MPC to date, and they are all classified as asteroids.

In December 2020, we observed asteroid 1998 KY$_{26}$ --- a potential Yarkovsky detection candidate \citep{Vokrouhlicky2000_yarko} and the target of the Hayabusa2 extended mission \citep{Hirabayashi2021} --- with the VLT (MPEC 2020-X181)\footnote{\url{https://www.minorplanetcenter.net/mpec/K20/K20XI1.html}} with the purpose of improving the measurement of the orbit.
At the same time, we also submitted two previously unreported observations from Mauna Kea in 2002 to the Minor Planet Center.
While the Yarkovsky effect was included in the fit and detected, the VLT observations showed a $-0.2''$ bias in declination that was larger than the astrometric uncertainties.
The bias was confirmed in additional observations from  La Palma (MPEC 2021-A42)\footnote{\url{https://www.minorplanetcenter.net/mpec/K21/K21A42.html}} and Mauna Kea (MPEC 2021-G127)\footnote{\url{https://www.minorplanetcenter.net/mpec/K21/K21GC7.html}}.
To ensure that the cause for this bias was not problematic data during the discovery apparition in 1998, we remeasured observations from the Modra Astronomical and Geophysical Observatory and Ond\v{r}ejov, which, together with radar \citep{Ostro1999}, were the only observations from 1998 we included in the fit. Despite our dataset revision, the bias still persisted.
Adding solar radiation pressure through the $A_1$ parameter was unfruitful, but adding an out-of-plane acceleration $A_3 (1\text{ au}/r)^2$ removed the bias in the 2020 data.

Subsequently, we observed 2005 VL$_1$ in 2021 (MPEC 2021-X95)\footnote{\url{https://www.minorplanetcenter.net/mpec/K21/K21X95.html}} and 2016 NJ$_{33}$ in 2022 (MPECs 2022-N88, 2022-O09)\footnote{\url{https://www.minorplanetcenter.net/mpec/K22/K22N88.html}}\footnote{\url{https://www.minorplanetcenter.net/mpec/K22/K22O09.html}} as Yarkovsky candidates. These observations were obtained with the 4.3 m Lowell Discovery Telescope in Arizona. Again, the collected data revealed a bias that could only be removed by adding an out-of-plane acceleration. 

Another case arose when we noticed the similarity between the orbits of 2010 VL$_{65}$\footnote{\url{https://ssd.jpl.nasa.gov/tools/sbdb_lookup.html\#/?sstr=2010vl65}} and 2021 UA12\footnote{\url{https://ssd.jpl.nasa.gov/tools/sbdb_lookup.html\#/?sstr=2021ua12}}.
The objects can be readily linked by fitting both datasets together, but only if $A_3$ is estimated as part of the fit. 

The last case is 2010 RF$_{12}$, which we observed in 2022 from Mauna Kea with the Canada France Hawaii Telescope and from Cerro Paranal with the VLT with the purpose of ruling out impact solutions detected by impact monitoring systems such as Sentry\footnote{\url{https://cneos.jpl.nasa.gov/sentry/}} and ESA's own system\footnote{\url{https://neo.ssa.esa.int/risk-list}}.
The observations we collected, in combination with an archival Mauna Kea set of images from 2011 taken with the Subaru Telescope (MPEC 2022-S77)\footnote{\url{https://www.minorplanetcenter.net/mpec/K22/K22S77.html}} once again revealed a bias that could only be removed by estimating $A_3$ while fitting the data.

Given these results, we decided to review the fit for 2006 RH$_{120}$ and found that in that case, there is also significant detection of an out-of-plane acceleration in addition to the radial and transverse ones.

\begin{figure*}
\begin{center}
  \includegraphics[scale=0.44,angle=0]{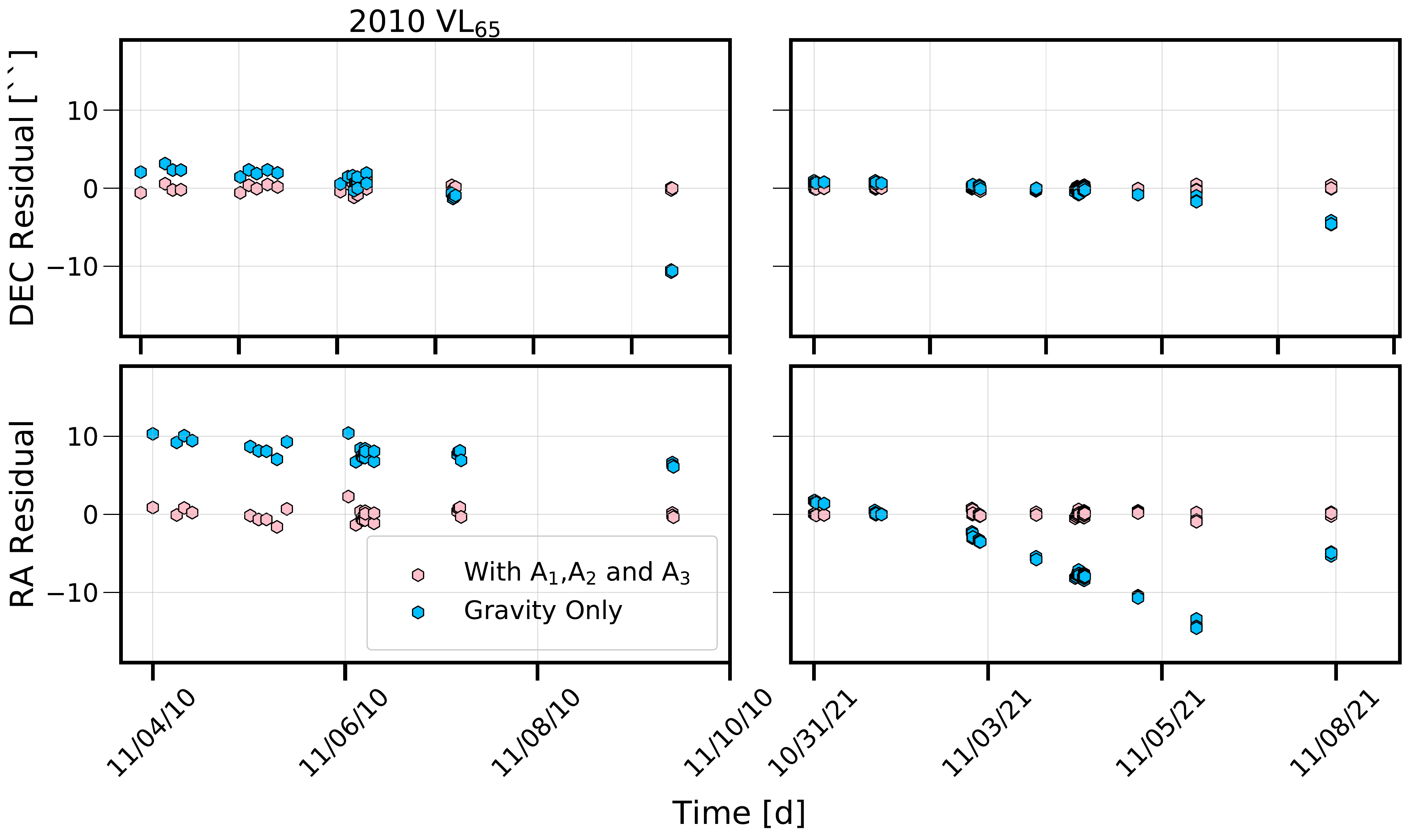}
   \includegraphics[scale=0.44,angle=0]{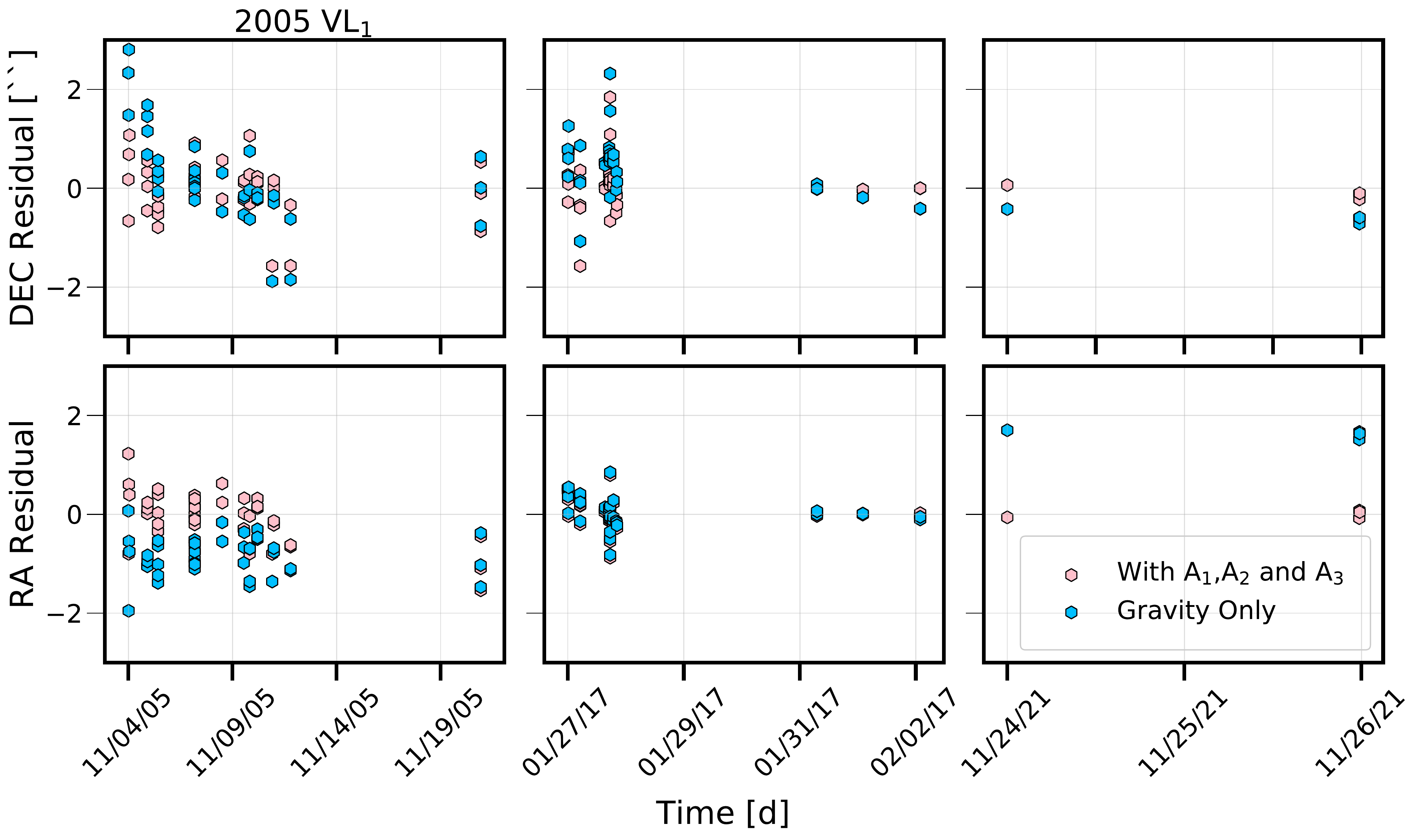}
    \caption{ The (signed) residuals of the astrometric fits (in arcseconds) for the right ascension and declination of 2010 VL$_{65}$ (top) and 2005 VL$_1$ (bottom) with (pink points) and without (blue points) nongravitational accelerations. The  gravity-only solution does not allow for outlier rejection in the astrometric fits. } 
 \label{fig:residuals} 
\end{center}
\end{figure*}

In Table \ref{table:objects}, we list each component, $A_1$, $A_2$, and $A_3$ of the best-fit nongravitational accelerations and their associated uncertainties. In order to avoid underestimating the uncertainty in the $A_1$, $A_2$, and $A_3$ parameters, we estimated all three parameters even when some of them (especially $A_1$) are not significantly detected. For each of these fits, we adopted $g(r) = (1\text{ au}/r)^2$. As examples, we show the
right ascension and declination residuals of the astrometric fits   with and without nongravitational accelerations for 2010 VL$_{65}$ and 2005 VL$_1$ in Figure \ref{fig:residuals}.  The improvement in the fit is evident from the figures. For VL$_{65}$, the gravity-only fit results in residuals as large as 10 arcsec. For 2005 VL$_1$, it is clear how the bias in 2021 (but also systematic errors in 2005) is removed by adding nongravitational accelerations.

In Figure \ref{fig:accelerations}, we show the best-fit $A_2$ and $A_3$ parameter for the non-radial nongravitational acceleration for each object. We compare the estimated $A_2$ and $A_3$ to the measured nongravitational accelerations in asteroids. We show both $A_1$, which is related to solar radiation pressure, and $A_2$, which is related to the Yarkovsky effect. Nominal by-eye fits of the acceleration, being inversely proportional to the nuclear radius computed using Equation \ref{eq:diameter_mag}, are shown.  There is a linear correlation between $\log |A_2|$  and $H$ (blue dashed line), as is expected from the Yarkovsky relation \citep{Vokrouhlicky2015_ast4}. The scatter around the mean is presumably due to different albedo, density, obliquity, or surface thermal inertia values. As a rule of thumb, the $\log |A_1|$ vs $H$ (pink dashed line)  again shows a linear trend with the same slope but an approximately order of magnitude larger normalization \citep{Vokrouhlicky2000}.  The non-radial component of the detected accelerations appears  to be more consistent with the extrapolation of the $A_1$ values. However, solar radiation pressure is mostly radial, and only a minor fraction  would project in the out-of-plane direction \citep{Vokrouhlicky2000}. Specifically, \citet{Vokrouhlicky2000} demonstrated that small off-radial  components of nongravitational accelerations could emerge from radiation pressure because  of variable albedo or non-spherical shapes. This suggests that these accelerations are inconsistent with radiation effects. 

Based on this analysis, we conclude that there are significant out-of-plane accelerations for all of these objects. If the out-of-plane accelerations in these objects are in fact caused by outgassing, it is possible that these are manifestations of nearly-polar jets with spin axis orthogonal to the orbit plane. 

The radial $A_1$ values are all consistent with zero acceleration, with the possible exception of 2016 NJ$_{33}$. Moreover, the $A_2$ values are much smaller in magnitude than the $A_3$ values and are generally consistent with the Yarkovsky effect. In the case of 2003 RM and 2006 RH120,  the $A_2$ values are anomalously high and presumably due to outgassing \citep{Chesley2016}.  Therefore, in the rest of the paper we focus on $A_3$ only. We note, however, that if high latitude jets are in fact causing the nongravitational accelerations, this would be somewhat distinct from typical comets that exhibit outgassing events isotropically over their surfaces. Moreover, incident stellar irradiation and subsequent surface temperature variations should not only affect the out-of-plane acceleration. This analysis can be generalized to larger accelerations if $A_1$ were to be statistically detected with longer data arcs. 

\begin{figure}
\begin{center}
       \includegraphics[scale=0.4,angle=0]{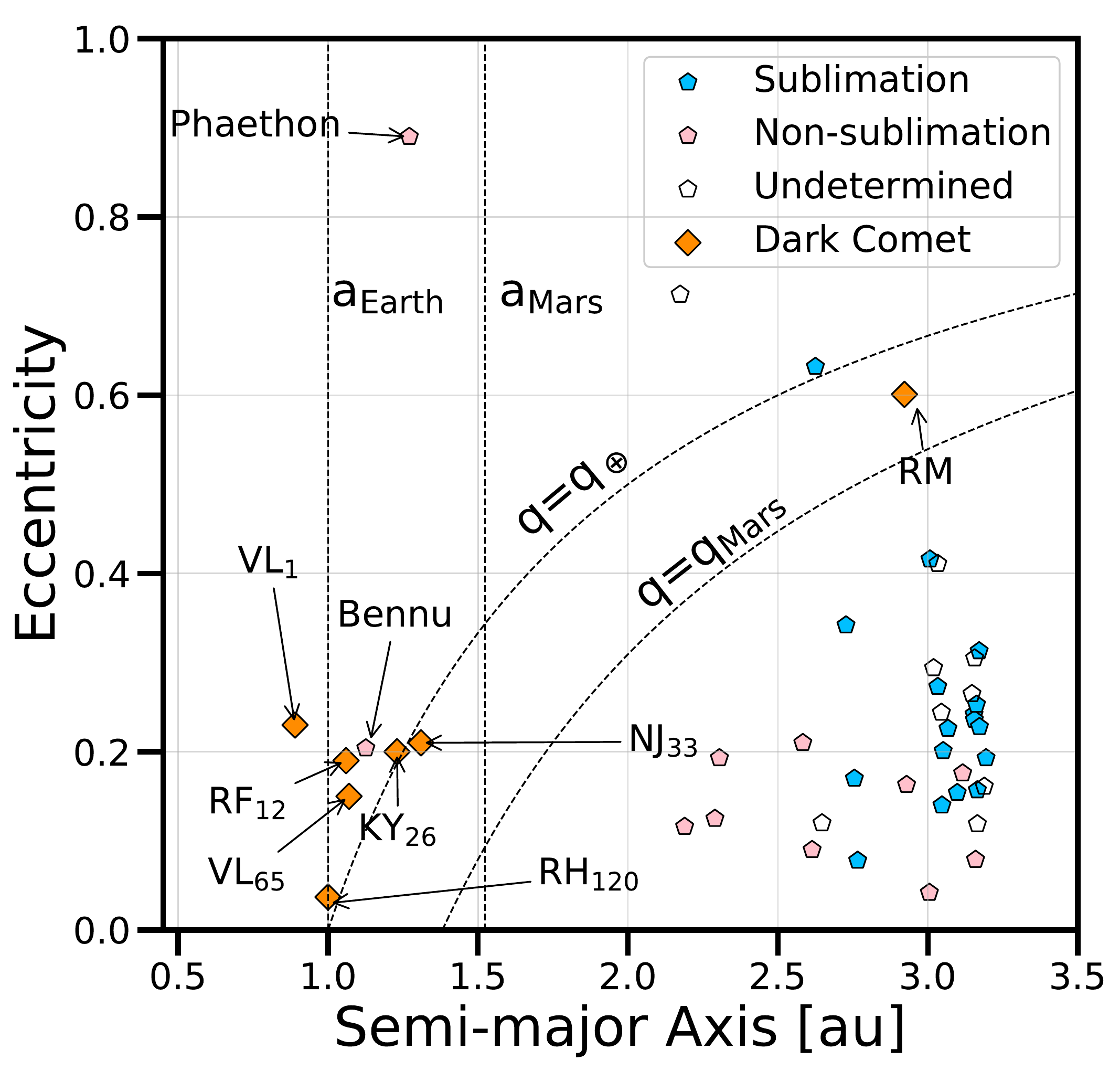}
    \caption{The location in semi-major axis and eccentricity space of the currently known active asteroids (pentagons, Table \ref{table:acive_asteroids}) and candidate dark comets (diamonds Table \ref{table:objects}). Active asteroids are color coded based on their activity sources: sublimation driven activity (blue), non-sublimation driven activity (pink), and unknown activity source (unfilled). The semi-major axis of Mars and the Earth and orbits with perihelia equal to that of Mars and the Earth are indicated with dashed lines.} 
 \label{fig:active_asteroids} 
\end{center}
\end{figure}

Table  \ref{table:objects} summarizes the observed orbital properties, absolute magnitudes and rotational periods (where measured) for each of these objects.    We use the following equation \citep{Pravec2007} to estimate the sizes for all objects except 1998 KY$_{26}$, 2003 RM and 2006 RH$_{120}$

\begin{equation}\label{eq:diameter_mag}
    2 R_{\rm Nuc} =\,\bigg(\, \frac{1329}{\sqrt{p}} \,\bigg)\, 10^{-0.2 \, H}\,,
\end{equation}
where $H$ is the absolute magnitude, $p$ is the geometric albedo and $R_{\rm Nuc}$ is the radius in kilometers. In Table  \ref{table:objects}, we assume $p=0.1$ to estimate the size of objects where this value has not been previously measured. The trajectories of the objects are unremarkable. In Figure \ref{fig:active_asteroids}, we show the location of these candidate ``dark comets''  as well as the location of all of the currently known active asteroids, whose properties are listed in Table \ref{table:acive_asteroids}. 

\subsection{Lack of Cometary Activity Detection}

\begin{figure}
\begin{center}
       \includegraphics[scale=0.36,angle=0]{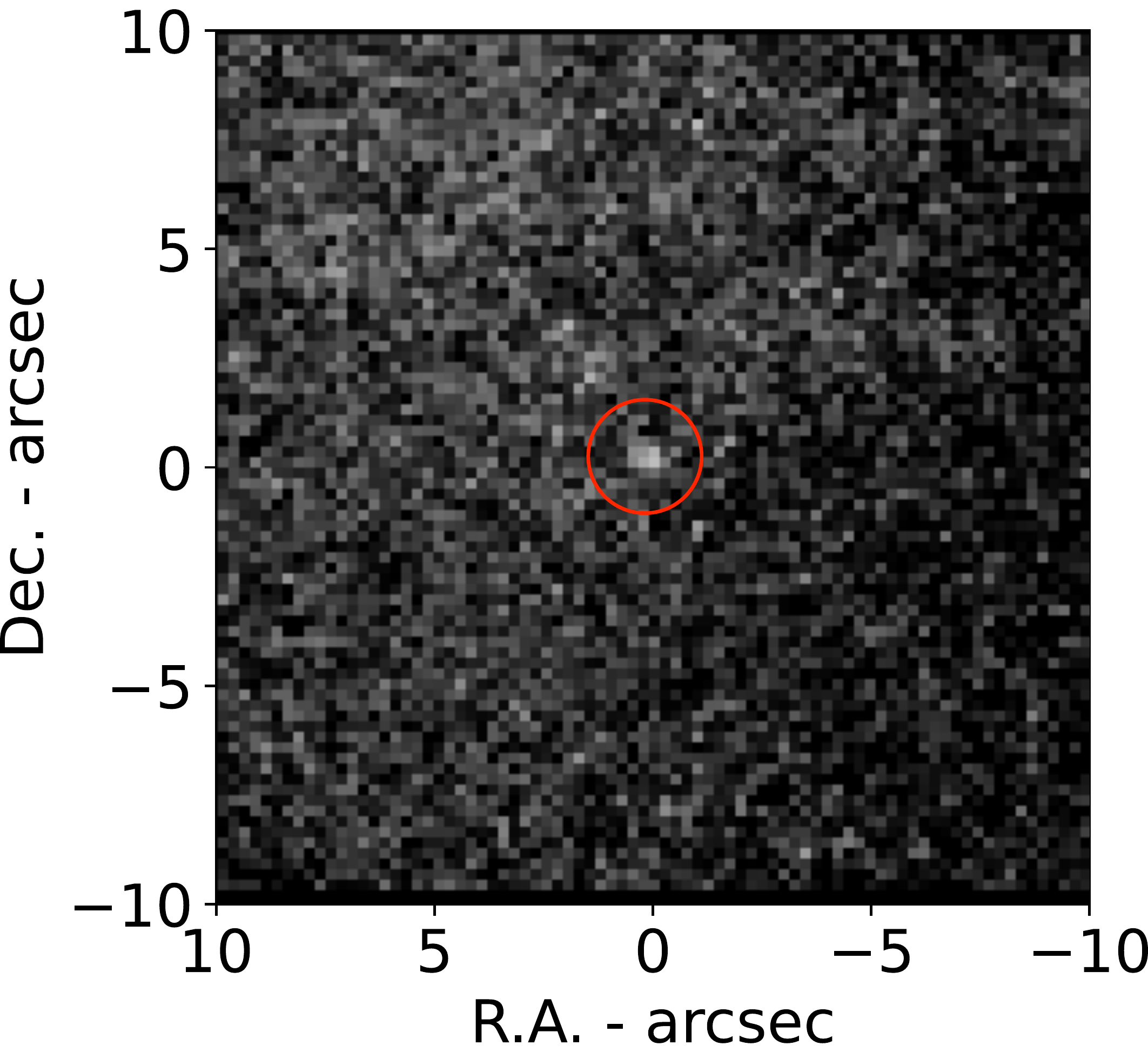}
    \caption{A stacked and cleaned image of 1998 KY$_{26}$ with VLT which spans 2 sets of  60 exposures, each of which is 30~s. The resulting temporal coverage is $\sim$ 1 h in Dec 2020. The image is 64$\times$64".This is a linear greyscale from 0 to 4 adu/s/pix. 1998 KY$_{26}$ is faint but visible in the final image and indicated with the red circle. There is no evidence for an extended cometary tail. } 
 \label{fig:ky26} 
\end{center}
\end{figure}

None of the objects studied in this paper have reported cometary activity. Out of all of our candidate targets, 1998 KY$_{26}$ was the only one for which we had meaningful data that could provide limits on the dust production.   The 1998~KY$_{26}$ images were obtained using the using the Unit Telescope~1 of the ESO Very Large Telescope (VLT) on Mount Paranal, Chile, with the FOcal Reducer and low dispersion Spectrograph 2 \citep[FORS2]{fors94}. The instrument was used without filter (white light), in order to reach the deepest possible limiting magnitude on asteroids and dust. FORS2 was equipped with  a mosaic of two MIT/LL CCD detectors, of which only the main chip, A, or CCID20-14-5-3, was used. The CCD was read with a 2$\times$2 binning, resulting in a $0.252''$ on sky. The de-biased and flat-fielded images were aligned on the position of the target. In Figure \ref{fig:ky26}, we show a stack of the images from December 2020 where no background object was too close from the object. This image contains 120 exposures of 30~s each, resulting in a 3600~s exposure.

\begin{figure}
\begin{center}
       \includegraphics[scale=0.52,angle=0]{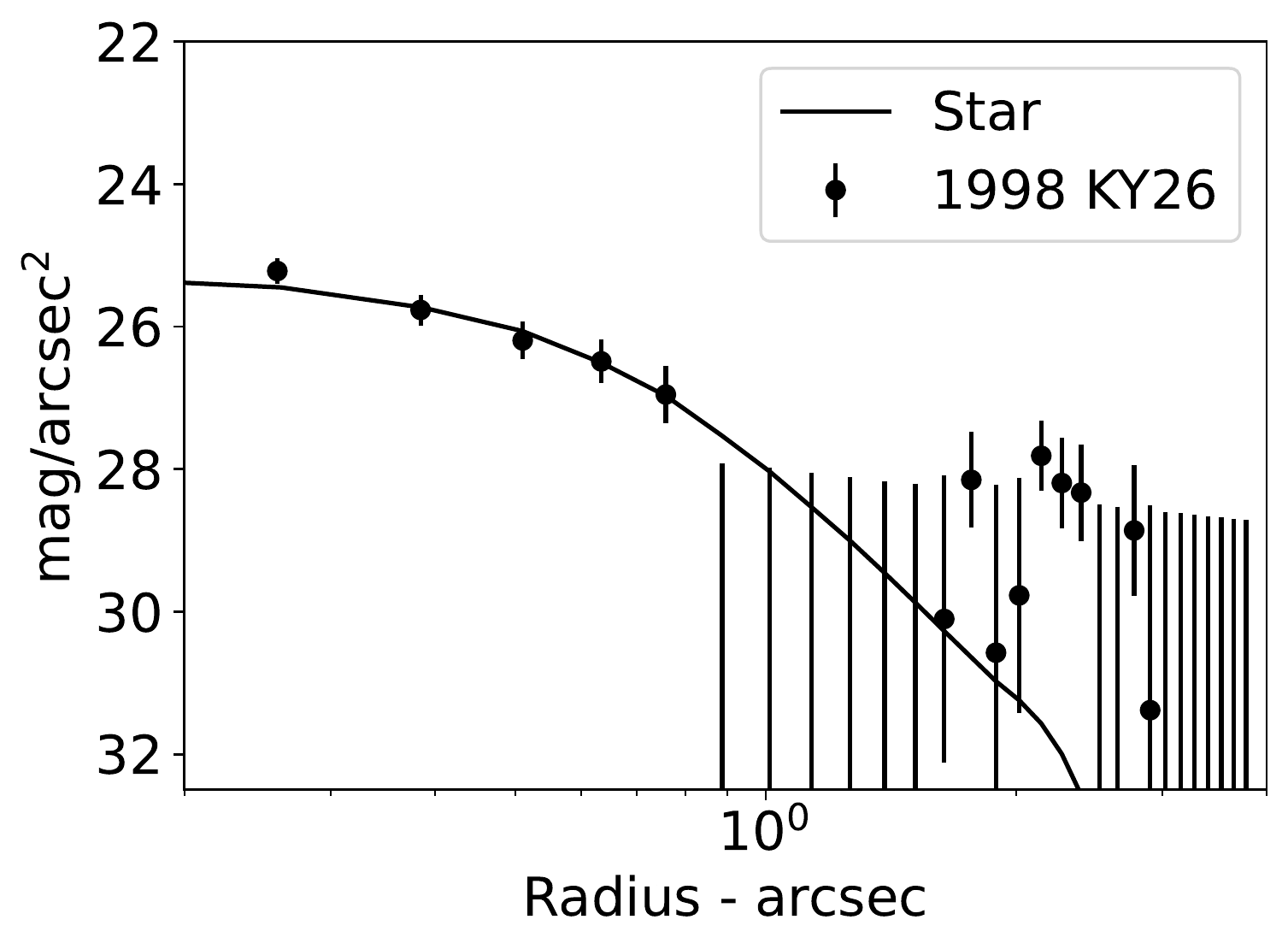}
    \caption{A standard profile analysis for dust activity in the deep stacked and cleaned image with VLT of  1998 KY$_{26}$ shown in Figure \ref{fig:ky26}. The   upper limit of dust present in the image is $\sim3$ kg. The solid line indicates the profile of our chosen field star scaled to the brightness of the object. The solid points indicate the integrated flux in concentric rings centered on the target. This analysis assumes that  the dust is composed of 1 micron sized dust grains, with an albedo $p=0.2$ and density $\rho=3000$ kg m$^{3}$, with $\sim1$ order of magnitude uncertainty. The x-axis increases to the right. } 
 \label{fig:dust_ky26} 
\end{center}
\end{figure}

The object was rather faint (26.1 for the first set, 25.5 for the second), requiring that the astrometric measurements be performed on stacks. The S/N in the individual images is barely above 1. This implies that the photometry would also require stacks. The object is the blob at the centre, outlined with a red circle. Although the object is very faint, the stacked images produced a solid detection with a combined S/N$\sim 12$. Furthermore, 1998 KY$_{26}$ is clearly visible in various sub-stacks, confirming without doubt the reality of the image and its identification with the object (the image is 64$"\times64"$). There is no evidence for an extended tail in the deep image; the variations observed in the sky around the object can be identified with residual of the various background objects passing behind the object. In order to further quantify the lack of dust around the target, we show in Figure \ref{fig:dust_ky26} its photometric profile.

The selection of the images composing the stack used to generate the profile ensured that no background object would contaminate the area of the profile. We obtained it by integrating the instrumental fluxes converted to surface brightnesses in a set of concentric circular apertures centered on the object. We use a conversion to magnitudes with a zero point of 27.8 \citep[from][]{Hainaut+21}. To scale the brightness of the object for comparison, we used the profile of a field star. This analysis does not yield a stringent constraint  because the noise and the background dominate very quickly. Because this object is so faint, the quantity of dust that could be undetected in the image is not  well constrained.  We estimate a coma mass  upper limit of $\sim 3$ kg of dust by integrating the corresponding flux  from 0.5$''$ to $2''$  (assuming 1 micron sized dust grains, with an albedo $p=0.2$, density $\rho=3000$ kg m$^{3}$). This corresponds to a magnitude for the dust of 24.0, which would be a 5$\sigma$ detection from 0.5 to 2" from the object. The uncertainty is $\sim 1$ order of magnitude because of all the assumptions required to convert from the magnitudes to the mass of dust (grain size, albedo, density). The corresponding upper limit on the dust production rate is  $\dot{M}_{\rm Dust}<0.2$ g s$^{-1}$. However, as discussed in the next section, outgassing could still explain the acceleration even if the upper limit was orders of magnitude higher.

\section{Implied Production Rates}\label{sec:production}
Based on the results presented in the previous section, it is clear that no mass loss has been detected in any of these dark comet candidates. However, the magnitudes of the nongravitational accelerations are inconsistent with being caused by the Yarkovsky effect or radiation pressure. Therefore, for the remainder of this paper, we hypothesize that the non-radial nongravitational accelerations are caused by outgassing. We calculate the implied production rates of H$_2$O using only the dominant, nonradial accelerations.

The production rate $Q(X)$ for a given species denoted by $X$ may be calculated using

\begin{equation}\label{eq:production_total}
    Q(X) = \,\bigg(\,\frac{M_{\rm Nuc} }{m_{\rm X}}\,\bigg)\,\,\bigg(\,\frac{ |A_i|}{v_{\rm Gas} \zeta}\,\bigg)\,.
    \end{equation}
In Equation \ref{eq:production_total}, $M_{\rm Nuc}$ is the mass of the nucleus, and  $m_{\rm X}$ and $v_{\rm Gas}$ are the mass and velocity of the  outgassing species. $A_i$ is the dominant component of the nongravitational acceleration used in the calculation. The variable $\zeta$ indicates the isotropy of the outflow, where  $\zeta$ = 1 corresponds to a  collimated outflow and $\zeta$ = 0.5 corresponds to an  isotropic hemispherical outflow.  The velocity of the gas can be related to the temperature  of the outflow, $T_\mathrm{Gas}$, using

\begin{equation}\label{eq:vgas}
    v_{\rm Gas} = \,\bigg(\, \frac{8k_B T_\mathrm{Gas}}{\pi m_\mathrm{X}}\,\bigg)^{1/2}\,.
\end{equation}

We calculate the implied production rates of H$_2$O outgassing using Equations \ref{eq:production_total} and \ref{eq:vgas} at perihelion (Table \ref{table:objects}).  We assume that  $\zeta=1$ and that the nuclei of the objects are spherical with radii listed in Table \ref{table:objects}, which in turn are calculated assuming an albedo of 0.1 and a bulk density of $\rho_{\rm Bulk} = 1\, {\rm g\, cm^{-3}}$. These production rates, $\sim10^{21}$ molec s$^{-1}$, are several orders of magnitude lower than typical cometary production rates, which are on the order of $\sim10^{26}$ molec s$^{-1}$ \citep[see Tables 1-3 in][]{Pinto2022}. The implied mass production of H$_2$O of 1998 KY$_{26}$ is $\sim0.03$ g s$^{-1}$, compared to $\lesssim1$ kg s$^{-1}$ for MBCs and $\sim10^2-10^3$ kg s$^{-1}$ for comets \citep{Jewitt2012,Jewitt22_asteroid}. If 1998 KY$_{26}$ has been outgassing continuously at this rate since its discovery, the total mass lost would be $\sim10^7$ g, $\sim0.1\%$ of its total mass. The analogous calculations for the other dark comet candidates yield similarly small mass loss fractions, implying that outgassing is an allowable mechanism. The production of MBCs has not led to detectable gas production \citep{Snodgrass2017}, except for a tentative detection of one candidate by \citet{Ferellec2022}. 

It should be noted that the case of 2016 NJ$_{33}$ requires a $\sim3\sigma$ significant $A_1$ radial acceleration to fit the astrometric data. We calculate the implied production rate using only the $A_3$ acceleration. However, if the $A_1$ is as large as the fit suggests, then the implied production of H$_2$O would be slightly higher. 

The characteristic mass-loss timescale $\tau_M$ for which this level of activity is sustainable is approximately given by 
\begin{equation}\label{eq:timescale}
 \tau_M \simeq  M_{\rm Nuc} \bigg/ \,\bigg(\,\frac{d M_{\rm Nuc}}{dt} \,\bigg)\, \simeq  \,\bigg(\,\frac{v_{\rm Gas} \zeta}{|A_i|} \,\bigg)\,
\end{equation}
A scaled relationship version of Equation \ref{eq:timescale}  gives

\begin{equation}\label{eq:masslosstime}
     \tau_M = 2.8\times10^4\, {\rm yr} \, \,\bigg(\,\frac{v_{\rm Gas}}{350\, {\rm m/s}} \,\bigg)\, \,\bigg(\,\frac{2 \times 10^{-11} {\rm \,au/day}^2  }{|A_i|} \,\bigg)\,.
\end{equation}

The orbits in our sample are similar to those of typical  evolved NEOs and may have had long residency times.  From Equation \ref{eq:masslosstime}, it is evident that these objects could not have been outgassing at these inferred rates for longer than $\sim10^4$ yrs. Therefore, if these objects are outgassing, then they  either (i) were not outgassing at this rate in the recent past or (ii) were only recently emplaced on these orbits. Objects with the trajectories similar to those of our sample dark comet candidates are thought to evolve from the main belt. Specifically, there is currently no known dynamical pathway to inject these objects from the JFC or LPC populations \citep[see, e.g., Figures 4-6 in][]{Granvik2018}. Moreover, these objects all have the highest probability of being injected onto their current trajectories  from the $\nu_6$ region in the innermost part of the asteroid belt. If our interpretation of the nongravitational acceleration based on outgassing is correct, it might have interesting population-wise implications on very small objects in the inner main belt. Although outside of the scope of this paper, future investigation of the dynamical history of these objects would be informative.

\section{Possible Explanation For Lack Of Visible Coma}\label{sec:nucelus_size}

\subsection{Cohesive Versus Rotational Forces For Surface Dust Retention}\label{subsec:rotation}

The objects in our sample  are between $\sim3-15$ m in radius, smaller than typical cometary nuclei that have been measured \citep{Jewitt2022}. Because small objects tend to rotate faster,  surface dust will experience stronger rotational forces. However, when considering typical cohesive forces on the surface regoliths of asteroids, it is evident that the rapid rotation of these objects is not sufficient to explain the lack of comae. 

Consider a particle at the equator and on the surface of a spherical nucleus with a radius  $R_{\rm Nuc}$ and a uniform density $\rho_{\rm Bulk}$.   In the absence of cohesive surface forces, the particle will be removed if its velocity  is greater than the escape velocity, $v_{\rm Esc}$, which is given by
\begin{equation}
    v_{\rm Esc}\,=\, R_{\rm Nuc} \, \sqrt{\,\frac{8\pi}{3} \, G \rho_{\rm Bulk}\,}\,.
\end{equation}

A scaled version of the same equation gives the following relationship:

\begin{equation}
    v_{\rm Esc} \, =\, 7.48 \,{\rm cm/s} \, \bigg(\,\frac{R_{\rm Nuc} }{1 \,{\rm km}}\,\bigg)\, \, \bigg(\,\frac{\rho_{\rm Bulk} }{1.0\, {\rm g\, cm^{-3}}}\,\bigg)^{1/2}\,.
\end{equation}

\begin{figure}
\begin{center}
       \includegraphics[scale=0.35,angle=0]{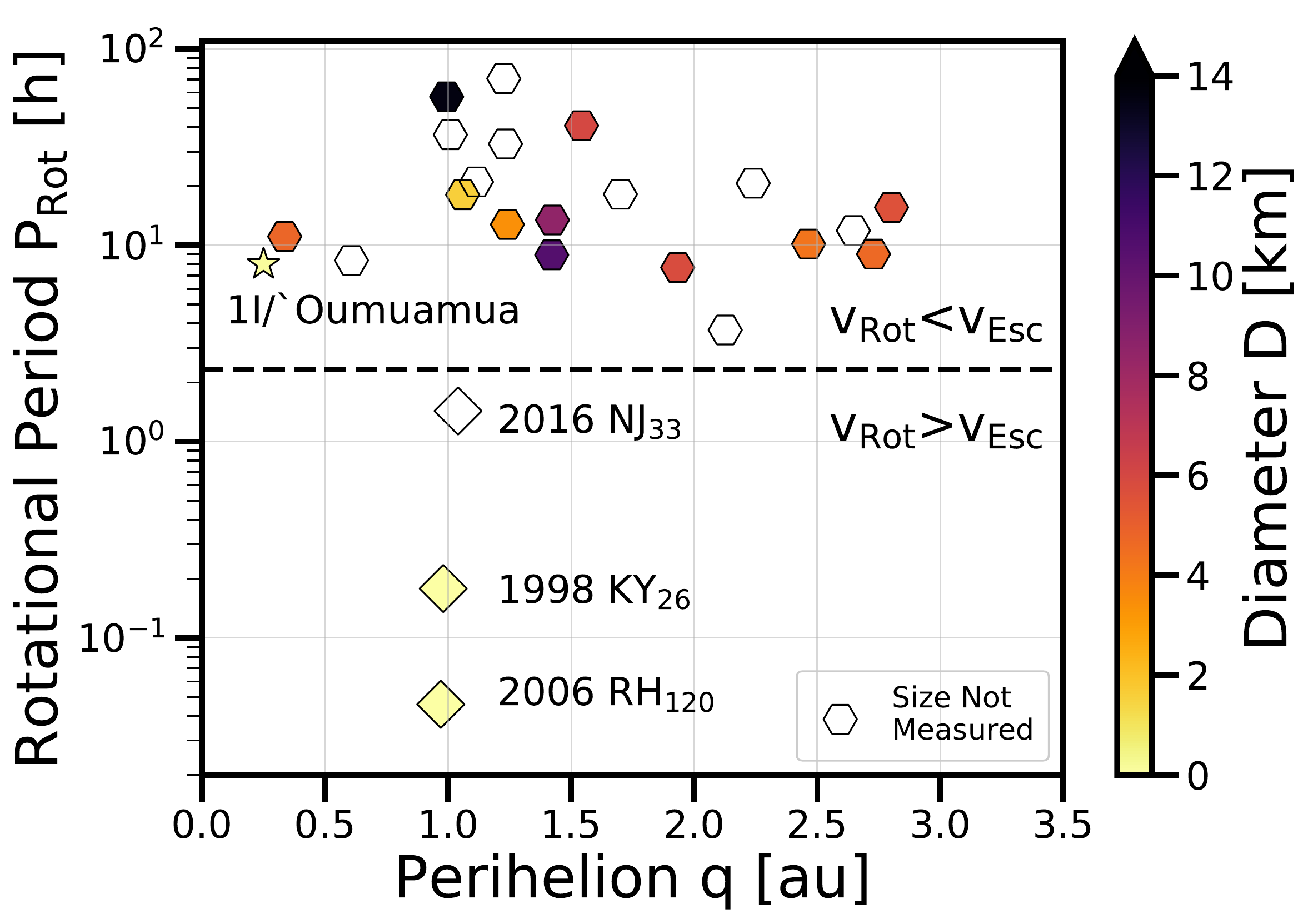}
    \caption{ Nuclear rotational periods of comets for which this values has been measured versus their perihelia. The nuclear diameter is shown in color for objects where this quantity has been measured. The location of 1998 KY$_{26}$, 2006 RH$_{120}$ and 1I/`Oumuamua are indicated, as well as the upper estimated rotation period of 2016 NJ$_{33}$. The dashed line shows the critical rotation period where self gravity balances with rotational forces. Data for each comet is listed in Table \ref{table:comets}.}
 \label{fig:rotate_diameter} 
\end{center}
\end{figure}

The rotational velocity $v_{\rm Rot}$ at the surface of a spherical nucleus rotating with rotational period $P_{\rm Rot}$  is   $v_{\rm Rot} = 2 \pi R_{\rm Nuc}/P_{\rm Rot}$. If the rotational velocity is greater than the escape velocity --- if $v_{\rm Rot}>v_{\rm Esc}$ --- then cohesionless particles will not remain on the surface. It is critical to note that this criteria only holds at the equator, and is diminished at higher latitudes. This criterion reduces to the following expression,

\begin{equation}
 P_{\rm Rot}\,<\,\sqrt{\,\frac{3 \pi   }{2 \, G \rho_{\rm Bulk}}}\,.
\end{equation}
The critical rotational period, $P_{\rm Crit}$ --- analogous to the rotation limit for a cohesion-less rubble-pile object \citep{Harris1979,Harris1979b,Harris1996,warner2009database} --- is given by

\begin{equation}\label{eq:critical}
    P_{\rm Crit} = 2.33 \, {\rm h} \, \bigg(\,\frac{1 \,{\rm g\, cm^{-3}}}{\rho_{\rm Bulk}}\,\bigg)^{1/2}\,\,.
\end{equation}
If an object has a rotational period less than the critical period, $P_{\rm Rot}<P_{\rm Crit} $, particles without cohesive forces will not be retained on the surface.  It has been established that the rotation periods of comets (when measured) are slower than those of asteroids \citep{Binzel1992,Jewitt2021,Jewitt22_asteroid}. In Figure \ref{fig:rotate_diameter} we show the rotational period vs perihelion for all comets for which the rotational period of the nucleus has been measured.  The comets all have slow nuclear rotational periods and therefore weak rotational forces operating on their surfaces. Clearly, 1998 KY$_{26}$, 2006 RH$_{120}$ and
2016 NJ$_{33}$ are rotating faster than this critical period. The comet data shown in Figure \ref{fig:rotate_diameter} is listed in Table \ref{table:comets}, with references for the cometary rotation period measurements.

However, it is worth noting that smaller dust may be retained on the surface by cohesion given by molecular forces, or some electrostatic interaction \citep{Rozitis2014}. In fact --- and somewhat counterintuitively --- smaller objects tend to have \textit{more dominant} cohesive forces \citep{Scheeres2010,Sanchez2020}. The rotational force acting on the equatorial particle is $f_{\rm Rot}=m_{\rm D} R_{\rm Nuc}\omega_{\rm Nuc}$ (where $\omega_{\rm Nuc}$ is rotational frequency of the comet nucleus and $m_{\rm D}$ is the mass of the dust grain). The cohesive forces can be estimated as $f_{\rm Coh}=\sigma_{\rm C} \Delta A_{\rm D}$, where $\sigma_{\rm C}$ is the strength of the regolith and $\Delta A_{\rm D}$ is the surface area of the grain \citep[as in Section 2 in][]{Sanchez2020}. Typical values of  $\sigma_{\rm C}$ are $\sim0-100$ Pa \citep{Hirabayashi2014}. Ignoring the effects of self gravity, the cohesive forces will dominate over the rotational forces when the fraction $f_{\rm Coh}>f_{\rm Rot}$. Assuming that $m_{\rm D}\simeq \rho_{\rm D} R_{\rm D}^3$ and $\Delta A_{\rm D}\simeq  R_{\rm D}^2$ for grain size $R_{\rm D}$ and density $\rho_{\rm D}$, the ratio of cohesive to rotational forces can be written as:

\begin{equation}\label{eq:cohesive_rot}
\begin{split}
    \frac{f_{\rm Coh}}{f_{\rm Rot}} \, =\, 3283 \,\bigg(\, \frac{\sigma_{\rm C}}{10\, {\rm Pa}} \,\bigg)\,
     \,\bigg(\,\frac{1\, {\rm g\,cm^{-3}}}{\rho_{\rm D}} \,\bigg)\,\\
      \,\bigg(\, \frac{1\, {\rm mm}}{R_{\rm D}}\,\bigg)\,
       \,\bigg(\, \frac{1 \, {\rm km}}{R_{\rm Nuc}}\,\bigg)\,  \,\bigg(\,\frac{P_{\rm Rot}}{1\,{\rm hr}} \,\bigg)^2\,.
       \end{split}
\end{equation}

\begin{figure*}
\begin{center}
       \includegraphics[scale=0.5,angle=0]{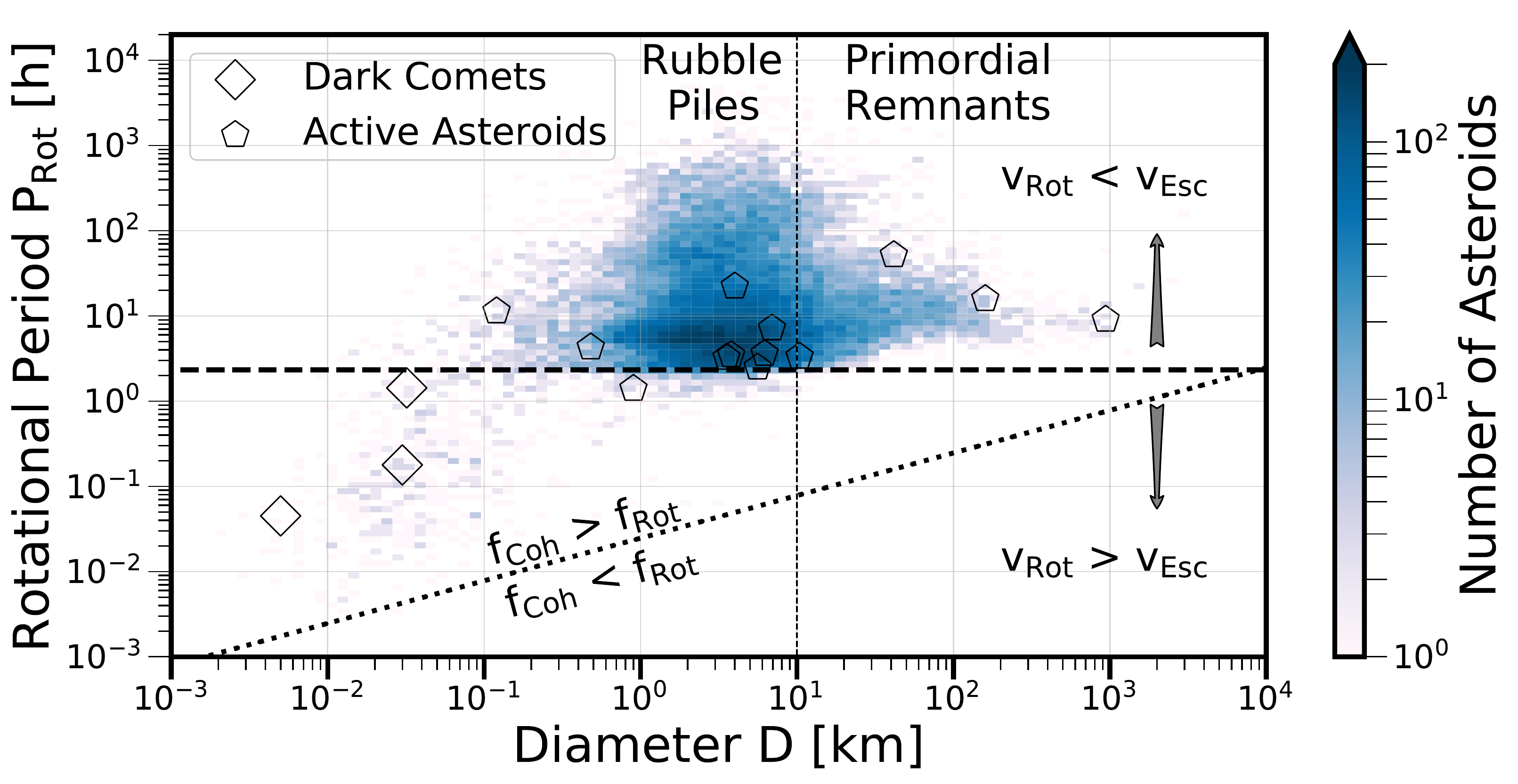}
    \caption{A 2-dimensional histogram showing the distribution of  rotational periods  vs.~diameters of asteroids. The critical period is shown in a dashed line, and the relationship where cohesive forces are equal to rotational forces (Equation \ref{eq:cohesive_rot} assuming $\sigma_{\rm C}=10$ Pa, $\rho_{\rm D}=$ 1 g cm$^{-3}$, $R_{\rm D}=1$ mm) is indicated with a dotted line.  These data are from  \cite{warner2009database}. Vertical dashed lines delineate size scales of rubble piles and primordial intact remnants. 1998 KY$_{26}$, 2016 NJ$_{33}$ and 2006 RH$_{120}$ are indicated with diamonds, while the other  candidate dark comets in our sample do not have rotation periods measured.  Active asteroids with sizes and rotational periods measured (Table \ref{table:acive_asteroids}) are indicated with pentagons.} 
 \label{fig:asteroid_rotate_diameter} 
\end{center}
\end{figure*}

In Figure \ref{fig:asteroid_rotate_diameter}, we show a 2-dimensional histogram of the diameter and rotational period of all asteroids for which these  properties have been measured. The diameter ranges for which asteroids are believed to be  rubble piles and primordial remnants as outlined in \citet{Walsh2018} are indicated.  We also show the location of all of the currently known active asteroids. It is evident that 1998 KY$_{26}$ and 2016 NJ$_{23}$ --- and presumably the remaining candidate dark comets --- are  distinct at the population level from active and inactive asteroids. The dashed  line shows the critical rotation period given by Equation \ref{eq:critical} and the dotted line shows where the cohesive forces equal the rotational forces computed with Equation \ref{eq:cohesive_rot}, assuming mm-sized dust grains, $\sigma_{\rm C}=10$ Pa, and $\rho_{\rm D}=1$ g cm$^{-3}$.  It is clear that although the candidate dark comets  rotate rapidly, typical cohesive forces will retain dust regolith on the surface. Therefore, the rapid rotation periods alone cannot explain the lack of detectable dust activity for these objects.

\subsection{Dependence of Spin-Up Timescale on Object Size}\label{subsection:spinup}

It is straightforward to show that, for a given sublimation torque, smaller objects  spin up faster (the same holds for the radiation induced torques, aka the YORP effect, which are however less important in the case of our sample of objects). Again, we consider a  spherical nucleus with radius $R_{\rm Nuc}$. If the outgassing induces average anisotropic rotational forces on the nucleus, $\vec{F}_{\rm Gas}$, the resulting torque is  $\vec{\tau}_{\rm Gas}$,

\begin{equation}
    \vec{ \tau}_{\rm Gas} \sim \vec{ R}_{\rm Nuc}  \times \vec{F}_{\rm Gas}\,.
\end{equation}

This torque produces a change of rotational frequency of the comet nucleus,  $\vec{\omega}_{\rm Nuc}$

\begin{equation}
   \vec{ \tau}_{\rm Gas} \sim M_{\rm Nuc} R_{\rm Nuc}^2 \frac{d \vec{\omega}_{\rm Nuc}}{dt}\,. 
\end{equation}
By combining the previous two equations, the rate of change of the spin of the nucleus is given by (neglecting the $\sin(\theta)$ term from the cross product)

\begin{equation}
    \frac{d \omega_{\rm Nuc}}{dt} \propto \,\frac{|F_{\rm Gas}| }{M_{\rm Nuc} R_{\rm Nuc} } \; .
\end{equation}
And, although this is a very general argument, surface forces on small bodies should be proportional to the surface area, i.e. $F_{\rm Gas}\propto R_{\rm Nuc}^2$. Thus, for a given bulk density, we obtain

\begin{equation}
    \frac{d \omega_{\rm Nuc}}{dt}\propto \frac{1}{ R_{\rm Nuc}^2}\,.
\end{equation}

Therefore, smaller  objects will be particularly susceptible to rapid spin up from outgassing torques.  In Figure \ref{fig:asteroid_rotate_diameter},  the most rapidly rotating objects are preferentially small ($\lesssim$~1 km) in diameter. It is important to note, however, that the magnitude of outgassing torques scales with the production, which could depend on the size of the nucleus with a different power than two assumed here. Based on the ratio of cohesive to rotational forces, it is possible that the rapid rotation of these small objects contributed to the removal of dust \textit{if} the objects were spun up via outgassing, and that same outgassing removed dust previously. However, the rotational forces alone, even for these rapidly rotating objecs, are not sufficiently strong to remove surface dust.

\subsection{Subsurface Dust Production}\label{subsection:dust_liits}

Given that our candidate dark comets do not exhibit dust comae, if their accelerations are caused by outgassing then they must not have retained significant dust on their surfaces. It is not clear what mechanism removed the dust from of the surface of the nuclei, given that rotation is not sufficient to overcome typical cohesive forces. It is possible that previous outgassing activity removed surface dust, and that the preferential rapid rotation in these dark comets is a leftover signature of this process. Processes such as 
meteorite bombardment (as with (101955) Bennu) and  fatigue due to thermal cycling will replenish surface dust, so for the nuclei to remain bare, the removal rate of dust via outgassing must be larger than the generation rate. Moreover, the episodic nature of these mechanisms militates against detection. It is also possible that significant surface material was cleared rapidly at birth, and the present-day dust creation rate is low enough to avoid apparent comae detection.

If these small objects are unable to retain surface material,  the only possible source of dust activity is from subsurface dust, which is entrained during ice sublimation. Assuming the nongravitational acceleration is due to volatile outgassing, we  calculate the maximum amount of dust that could be produced via such entrainment, assuming a dust-to-gas mass ratio $\mathcal{Z}= 0.01$ typically assumed in the ISM and protoplanetary disks \citep{Birnstiel2010}.
The dust production rate $\dot{M}_{\rm Dust}$ for a given outgassing species $X$ can be calculated from the gas production rate (Table \ref{table:objects}) by multiplying the total gas mass production of the outflow by the dust-to-gas ratio. Assuming H$_2$O driven gas production, the dust production  can be calculated using

\begin{equation}\label{eq:dustprod}
\begin{split}
    \dot{M}_{\rm Dust}(X)  =3\times10^{-4} \, {\rm g \,s^{-1}}  
    \, \bigg(\, \frac{\mathcal{Z}  }{0.01} \, \bigg)\,\\\,
      \, \bigg(\,\frac{Q(X)}{10^{21} {\rm molec}\,  {\rm s}^{-1} }\, \bigg)\,
      \, \bigg(\, \frac{m_{\rm H_2O}}{m_X}\, \bigg)\,.
      \end{split}
\end{equation}

Due to the small gas  production rates required to power the accelerations of order $10^{19}-10^{21}$ molec s$^{-1}$, the corresponding dust production rates are extremely small.  For instance, in the case of H$_2$O outgassing from 1998 KY$_{26}$, the dust production rate would be $\sim10^{-4}$ g s$^{-1}$.  For context, (101955) Bennu exhibited $\dot{M}_{\rm Dust}\sim10^{-4}$ g s$^{-1}$ which was only measurable in situ \citep{Hergenrother2020}. By comparison, typical observable cometary dust production rates are $\gtrsim 10^4$ g s$^{-1}$ \citep{Delsemme1976}. As a point source analogue, micron sized dust production limits from deep stacked optical images of 1I/`Oumuamua were $\dot{M}_{\rm Dust} \le$ 0.2 -- 2 g s$^{-1}$ \citep{Jewitt2017,Meech2017}, $\gtrsim10^3\times$ greater than these dust production rates. Therefore, in the absence of surface dust, small, rapidly rotating, and weakly outgassing bodies should not produce detectable dust activity. It is not surprising that these objects have not been identified as active asteroids \citep{Jewitt22_asteroid}. 

In the absence of extended tails in images, activity may be detected via drastic brightness changes in light curves. For the case of (101955) Bennu, which displayed no tail, \citet{Hergenrother2020} estimated that  the dust production $\dot{M}_{\rm Dust}\sim10^{-4}$ g s$^{-1}$ would not be detectable in light curves. They estimated the detection threshold of ground based photometric surveys as  0.1 magnitudes, or 9.6$\%$ of its absolute magnitude.  Given (101955) Bennu's average projected surface area of $1.9 \times 10^9$ cm$^2$, this sensitivity limit  corresponds to the release of dust particles with total area $1.8 \times 10^8$ cm$^2$. The largest ejection event on  6 January produced a surface area of ${\sim}170$--$190$ cm$^2$. Therefore, even the largest production event recorded for (101955) Bennu would have been undetectable with ground based observations by $\sim10^6$ orders of magnitude.  And so, again, it is not surprising that activity has not been detected in these objects, since the implied dust production is comparable to that of (101955) Bennu. It is important to note that  (101955) Bennu does not have a significant nongravitational acceleration due to particle ejection, and the activity is thought to stem from other sources such as meteoroid bombardment with a granular surface prepped by thermal fracturing \citep{Chesley2020}.

We perform an analogous calculation for 1998 KY$_{26}$. The projected surface area of 1998 KY$_{26}$ with $R_{\rm Nuc}=15$~m is $\sim 7\times10^6$ cm$^2$. We assume that the dust produced had the same properties as the particles ejected from (101955) Bennu, with a resulting projected surface area of 170-190 cm$^2$. Because 1998 KY$_{26}$ is smaller than (101955) Bennu, the change in brightness would be $\sim2.5\times10^{-5}$  or $\sim0.0025\%$ of its absolute magnitude. This is orders of magnitude below the uncertainty level (0.1 mags) of the photometric  measurements obtained during  the 1998 apparition \citep[Figure 3 in][]{Ostro1999}. For the remaining objects, extant photometric data (some of which is reported without error bars) is not sufficient to capture such brightening events (not shown). Although there has been no activity reported for any of our candidate dark comets, the  dust production levels estimated here are not detectable in extant data by orders of magnitude. Moreover,  even if the photometric measurements did have the absolute accuracy  required to detect this level of activity, the \textit{intrinsic} variation due to (i) rotation, (ii) phase variation, and (iii) changing of geometry would mask this relatively small activity signature. To detect this level of activity photometrically we would first require a high-precision shape model with high fidelity information regarding the scattering properties of the surface.

\begin{figure*}
\begin{center}
\includegraphics[scale=0.55,angle=0]{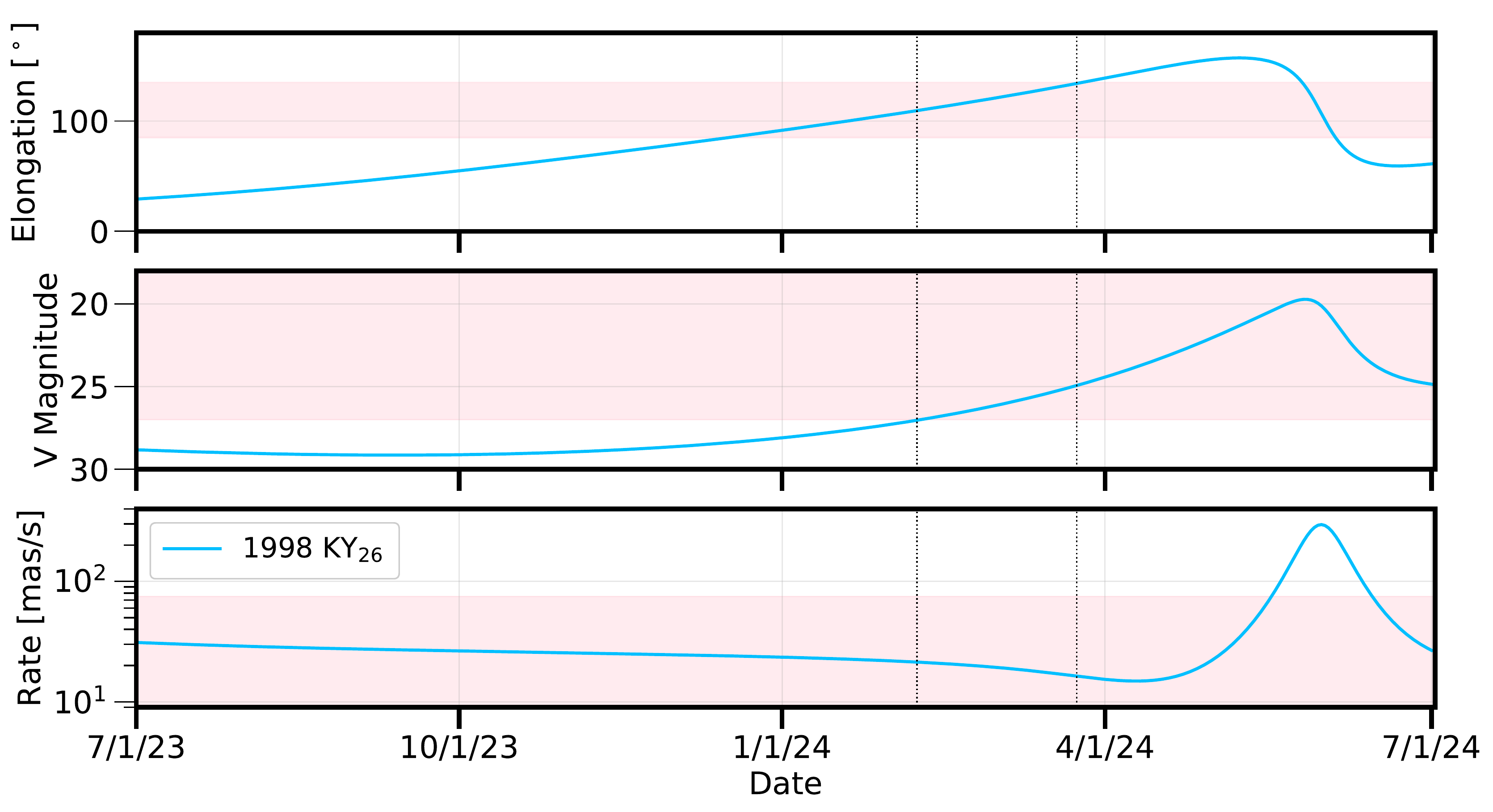}
\caption{The observability of the 1998 KY$_{26}$ with JWST from July 2023-2024. We show the Sun-Observer-Target apparent solar elongation angle as seen from the observers' location (top), the apparent visual magnitude (middle) and the rate of motion on the sky (bottom). The other candidate dark comets are not observable during this time period. The pink shaded region corresponds to nominal observability criteria for JWST --- specifically   $85^\circ<$ Elongation $< 135^\circ$, V-mag $< 27$ mag and Rate of motion $< 75$ mas s$^{-1}$. The vertical dotted lines indicate the time window between February and March 2024 when all three observability criteria will be satisifed.  }\label{Fig:observability}
\end{center}
\end{figure*}

\section{Follow-Up Observations}\label{sec:follow_up_observations}
The peculiar behavior of these objects compared to known comets and asteroids makes them prime targets for follow-up observations.   Given that volatile outgassing is a plausible mechanism to explain the out-of-plane nongravitational accelerations, spectroscopic observations are needed to detect or place limits on the presence of candidate volatiles (H$_2$O, CO$_2$, CO) around these objects.  Given the low production rates, the sensitivity of the \textit{James Webb Space Telescope} (JWST) will likely be required to obtain meaningful constraints.  Additionally, deep imaging is needed to improve the limits on dust activity levels. Moreover, additional astrometric observations would help to constrain the radial acceleration $A_1$, and may show that the nongravitational accelerations of these objects are higher than those reported in this paper. 

In Figure  \ref{Fig:observability}, we show the solar elongation, visual magnitude and rate of motion on the sky for 1998 KY$_{26}$.  Based on the observing constraints of JWST,  1998 KY$_{26}$ is the best candidate for observations in the near future (through 2024).  Improved constraints on the nature of this  objects will help to resolve whether they are indeed part of a previously undescribed class of Solar System bodies.

It was  announced in  September 2020 that Hayabusa2 will visit 1998 KY$_{26}$ after performing two swing-by maneuvers at Earth in the extended mission \citep{Hirabayashi2021}. The spacecraft is equipped \citep{Watanabe2017} with a suite of instruments, the first being an Optical Navigation Camera (ONC) with one telescopic  and two wide-angle  cameras with seven filters at 0.39 microns  (ul-band), 0.48 microns (b-band), 0.55 microns (v-band), 0.59 microns (Na), 0.70 microns (x-band), 0.86 microns (w-band), and 0.95 microns (p-band) \citep{Kameda2015,Kameda2017,Suzuki2018,Tatsumi2019}. The Thermal Infrared Imager (TIR) is capable of measuring surface roughness and thermal emmissivity and  inertia  via high resolution images covering 8-12 microns of thermal infrared emission which capture surface temperature ranges from 150-460 K \citep{Arai2017,Okada2017,Takita2017}. The Laser Altimeter (LIDAR) \citep{Mizuno2017,Senshu2017,Yamada2017}  can detect dust grains in the vicinity of an asteroid, with quoted sensitivity to asteroidal dust grains similar to the Hayabusa sample surrounding the target if the number density is $>10^{5}$ m$^{-3}$ \citep{Senshu2017}. It is also equipped with a Near-Infrared Spectrometer (NIRS3) capable of obtaining spectra at near infrared wavelengths of 1.8 to 3.2 microns, designed to measure reflectance spectra of absorption bands of hydrated and hydroxide minerals \citep{Iwata2017}. Therefore, the nature of the nongravitational acceleration of 1998 KY$_{26}$ should be identified definitively by Hayabusa2.

\section{Conclusions}\label{sec:conclusions}

In this paper, we identified five inactive objects lacking visible coma, which nonetheless showed significant nongravitational acceleration in the out of the plane direction. These objects are part of the NEO population, and are all characterized by non-peculiar orbits (e.g.~typical semi-major axes, eccentricity, and inclination). As a rule, these objects are small ($R_{\rm Nuc}\sim 3-16$~m) and have rapid rotation periods (when measured). 

While the nature of these objects remains uncertain, we showed that an outgassing mechanism can plausibly explain the out-of-plane non-gravitation acceleration without producing a visible dust coma. This is largely due to the small size of the bodies, which means that (i) relatively low gas production rates are needed to explain the accelerations, and (ii) the nuclei may not have surface dust possibly due to continual cleansing from outgassing.   For objects that have lost their surface dust, the only contribution to dust activity is the entrainment of dust from the subsurface during ice sublimation.  Based on the gas production rates inferred for the bodies, we show that the production of dust via entrainment is extremely small and well within the observational limits. It is possible that these objects are Solar System analogues to 1I/`Oumuamua, which also exhibited a nongravitational acceleration and lack of cometary activity.

Follow-up observations are critical to understanding the nature of these unusual objects, potentially members of a new class of ``dark comets'' in the Solar System.  In particular,   observations with JWST would reveal whether their nongravitational accelerations are in fact due to volatile outgassing (e.g. H$_2$O, CO$_2$, or CO), or whether a new mechanism is needed to explain their peculiar properties.

\section{Acknowledgments}
We thank Dave Jewitt, John Noonan, Masatoshi Hirabayashi, Dong Lai,  Fred Adams, Nikole Lewis,  Konstantin Batygin, Samantha Trumbo, Gregory Laughlin,  Juliette Becker, Mike Brown, Ngoc Truong, J.T. Laune, and Jonathan Lunine for useful conversations and suggestions. We thank Dave Jewitt and Henry Hsieh for compiling the data in Table 3 and for granting permission to reproduce it.  We thank the scientific editor Faith Vilas for securing constructive and rapid referee reports. We thank the two anonymous reviewers for insightful comments and constructive suggestions that strengthened the scientific content of this manuscript.

DZS acknowledges financial support from the National Science Foundation  Grant No. AST-17152, NASA Grant No. 80NSSC19K0444 and NASA Contract  NNX17AL71A from the NASA Goddard Spaceflight Center. Part of this research was carried out at the Jet Propulsion Laboratory, California Institute of Technology, under a contract with the National Aeronautics and Space Administration (80NM0018D0004).
DV was supported by the Czech Science Foundation, grant GA21-11058S. This work was partially supported by NASA grant number NNX17AH06G (PI N. Moskovitz) issued through the Near-Earth Object Observations program to the Mission Accessible Near-Earth Object Survey (MANOS). DJT acknowledges support from NASA Grant 80NSSC21K0807.  
The Arecibo planetary radar observation presented is this work was supported by the National Aeronautics and Space Administration's (NASA’s) Near-Earth Object Observations program through grant  No. NNX13AQ46G awarded to Universities Space Research Association (USRA). Reduction of the data was performed under grant Nos. 80NSSC18K1098 awarded to the University of Central Florida (UCF). The Arecibo Observatory is a facility of the National Science Foundation (NSF) operated under cooperative agreement by UCF, Yang Enterprises, Inc., and Universidad Ana G. M\'{e}ndez. This paper is partially based on observations collected at the European Southern Observatory under ESO programme 105.202E.002.

\bibliography{bibliography}{}
\bibliographystyle{aasjournal}

\appendix

\section{Rotation period of 2016 NJ$_{33}$}\label{sec:nj33_radar}

\begin{figure}
\begin{center}
       \includegraphics[scale=0.38,angle=0]{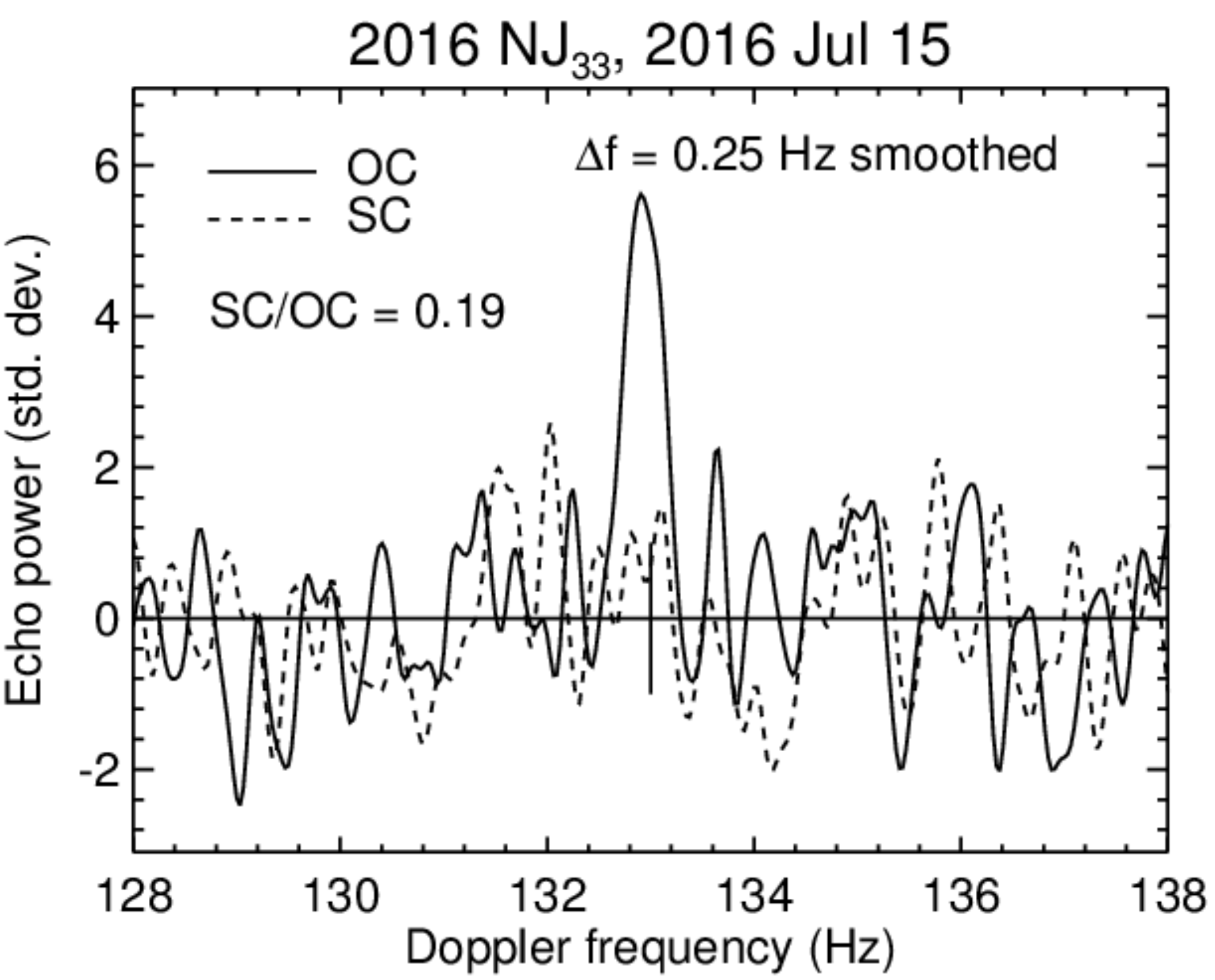}
    \caption{A radar CW spectrum of 2016 NJ$_{33}$ taken with the Arecibo Observatory.   The CW spectrum shows that the minimum rotation period should be between 0.41 and 1.99~h. Because we have no information regarding the pole orientation of the object, it is possible that the rotational period is much more rapid. The low SC/OC ratio exclude an E-type classification \citep{Benner_2008}. Reproduced from \url{https://www.naic.edu/~pradar/asteroids/2016NJ33/2016NJ33.2016Jul15.s0p25Hz.cw.gif}}
 \label{fig:NJ33} 
\end{center}
\end{figure}

For 2016 NJ$_{33}$, we estimate the rotational period using one continuous wave (CW) observation obtained with the Arecibo Observatory planetary radar system, shown in Figure \ref{fig:NJ33}. The CW spectrum shows a bandwidth of $0.50 \pm 0.02$ Hz. Using the relation linking the bandwidth of a CW spectrum and the rotation period (see \citet{Virkki2022} for more information on CW radar observation and their analysis) we estimate  that the rotation period of 2016 NJ$_{33}$ is between 0.41 and 1.99 h with a mean of 1.43 h with 3 sigma confidence. We obtained this result by simulating random pole orientations and a normal distribution of the bandwidth ($0.50 \pm 0.02$~ Hz) and the diameter (32 $\pm$ 3~m). This estimation of the rotational period should be treated with caution because (i) we only have one observation of the object, (ii) an observation very close to pole-on geometry would result in a very fast rotation period, (iii) we assume a spherical nucleus and if the object is highly elongated, a very different   cross-section  would be allowable. However,  there is a low probability of  randomly observing the object pole-on.

\section{Data for active asteroids and cometary nuclei}

Table \ref{table:objects} is reproduced from \citet{Jewitt22_asteroid}. In Table \ref{table:comets}, we show relevant data and references for comets with measured nuclei rotation rates and sizes. In Table \ref{table:acive_asteroids}, we list active asteroid properties.

\setlength{\tabcolsep}{1.5pt}
\begin{table*}
\begin{rotatetable*}
\begin{center}   
\caption{\textbf{Objects with Anomalous Nongravitational Accelerations}. The best-fit nongravitational acceleration parameters for each object in our sample, with formal (1-$\sigma$) uncertainties. The statistical significance is given for each acceleration and calculated as the ratio of the magnitude of the acceleration to its uncertainty. We also include the object 2003 RM, for which we report anomalously large nongravitational acceleration in the accompanying paper Farnocchia et al. (submitted 2022). Statistically robust results ($\sigma \geq3$) are found for the $A_3$ component of all newly reported objects in this paper. For the $A_2$ component, only 2016 NJ$_{33}$, 2003 RM and 2006 RH$_{120}$ have robust detections. We calculate implied production rates of H$_2$O for each object in our sample at perihelion. We calculate this using the most  significant component of the acceleration, and assuming an albedo of $p=0.1$ and that the molecules are released at a temperature of 100 K.}
\label{table:objects}
\footnotesize
\begin{tabular}{lccccccc l@{}c@{}lc l@{}c@{}lc l@{}c@{}lc c} 
 Object  & $a$ & $e$ & $i$ & $q$ & $H$ & $R_{\rm Nuc}$ & $P_{\rm Rot}$ & \multicolumn{3}{c}{$A_1$}  & Signif. & \multicolumn{3}{c}{$A_2$} & Signif. & \multicolumn{3}{c}{$A_3$}& Signif. & $Q$(H$_2$O)\\
 & [au] &  & [$^\circ$] & [au] & [mag] &[m]& [h] & \multicolumn{3}{c}{[$10^{-10}$ au d$^{-2}$]} & [$\sigma$]&   \multicolumn{3}{c}{[$10^{-10}$ au d$^{-2}$]} &[$\sigma$]&  \multicolumn{3}{c}{[$10^{-10}$ au d$^{-2}$]}  &[$\sigma$]& [10$^{20}$ molec s$^{-1}$]\\
 \hline
1998 KY$_{26}$ & 1.23 & 0.20 & 1.48 & 0.98 & 25.60& 15& 0.178       &\phantom{-}1.73 &$\pm$&0.91 & 2   & -0.00126&$\pm$&0.00061 &2    &\phantom{-}0.320  &$\pm$&0.115 &3    & 9.19\\
2005 VL$_1$ & 0.89 & 0.23 & 0.25 & 0.69 & 26.45 & 11 &              & -6.66 &$\pm$&8.02 &$<1$ & -0.00711&$\pm$&0.00592 &1    & -0.240 &$\pm$&0.041 &6    &5.5
\\
2016 NJ$_{33}$ & 1.31 & 0.21 & 6.64 & 1.04 & 25.49&16&$0.41$-$1.99$ &\phantom{-}9.28 &$\pm$&2.96 &3    & -0.00566&$\pm$&0.00193 &3    &\phantom{-}0.848  &$\pm$&0.163 &5    &24.8 \\
2010 VL$_{65}$ & 1.07 & 0.14 & 4.41 &0.91 & 29.22&3&                &\phantom{-}6.57 &$\pm$&13.0 &$<1$ & -0.00146&$\pm$&0.00534 &$<1$ & -0.913 &$\pm$&0.130 &7    & 0.24  \\ 
2010 RF$_{12}$&1.06 & 0.19  & 0.88  &0.86 &28.42 &4 &               &\phantom{-}0.488&$\pm$&0.597&$<1$ & -0.00136&$\pm$&0.00286 &$<1$ & -0.168 &$\pm$&0.021 &8    & 0.12 \\ 
2006 RH$_{120}$& 1.00 & 0.04  & 0.31  & 0.96 & 29.50 & 2-7 & 0.046 &\phantom{-}1.38 &$\pm$&0.08 &18   & -0.507  &$\pm$&0.0637  &8    & -0.130 &$\pm$&0.032 &4    & 1.5\\
2003 RM & 2.92 & 0.60 & 10.86 & 1.17 & 19.80 & 230 &                & -1.045&$\pm$&1.217&$<1$ &\phantom{-}0.0215 &$\pm$&0.0004  &56   &\phantom{-}0.0156&$\pm$&0.0543&$<1$ & 1600 \\
\hline
\end{tabular}
\end{center}
NOTES: The size and rotational period of 1998 KY$_{26}$ was first reported by \citet{Ostro1999}. The  rotational period of 2016 NJ$_{33}$ is estimated from radar CW spectral observations taken with the Arecibo Observatory planetary radar system (see Appendix ~\ref{sec:nj33_radar}). The nuclear radius and period of 2006 RH$_{120}$ were reported by \citet{Kwiatkowski2009}. The remaining objects do not have reported measurements of size or rotational period. We estimate the remaining sizes with Equation \ref{eq:diameter_mag} assuming a geometric albedo of $p=0.1$. 
\end{rotatetable*}
\end{table*}

\setlength{\tabcolsep}{6pt}

\begin{table*}
\centering
\caption{\textbf{Comets with nuclear rotation periods measured.} The data is drawn from JPL small body database.}
\begin{tabular}{ ccccccc } 
 Object Name & $q$ [au] & $i$ [$^\circ$] & Classification & Diameter [km] & $P_{\rm Rot}$ [h] & Reference\\
 \hline
     P/2006 HR30 (Siding Spring)&	1.226&	31.88&	HTC	&&	70.7&\\
     C/2001 OG108 (LONEOS)&	0.994&	80.25&	HTC	&13.6&	57.12&\citet{Abell2005} \\
     &	&	&	&&	&\citet{Fernandez2005} \\
    9P/Tempel 1&	1.542&	10.47&	JFC	&6&	40.7&\citet{Ahearn2005}\\
     P/2016 BA14 (PANSTARRS)&	1.009&	18.92&	JFC	&&	36.6&\citet{Warner2016}\\
  162P/Siding Spring&	1.233&	27.82&	JFC	&&	32.853&\citet{Kokotanekova2017}\\
  333P/LINEAR&	1.115&	131.88&	JFC	&&	21.04&\citet{Hicks2016}\\
   94P/Russell 4&	2.24&	6.18&	ETC		&&20.7&\citet{Kokotanekova2017}\\
   93P/Lovas 1&	1.7	&12.2&	JFC	&&	18.2&\citet{Kokotanekova2017}\\
  103P/Hartley 2&	1.059&	13.62&	JFC	&1.6&	18.1&\citet{Belton2013}\\
  &	&	&	&&	&\citet{Harmon2011} \\
   47P/Ashbrook-Jackson&	2.802&	13.05&	JFC	&5.6&	15.6&\citet{Kokotanekova2017}\\
   49P/Arend-Rigaux&	1.424&	19.05&	JFC &	8.48&	13.45&\citet{Eisner2017}\\
   67P/Churyumov-Gerasimenko&	1.241&	7.05&	JFC	&3.4&	12.76129&\citet{Mottola2014}\\
   &	&	&	&&	&\citet{Sierks2015} \\
  149P/Mueller 4&	2.647&	29.75&	JFC	&&	11.88&\citet{Kokotanekova2017}\\
    2P/Encke&	0.336&	11.78	&ETC &	4.8	&11.083&\citet{Fernandez2005b}\\
 &	&	&	&&	&\citet{Lowry2007} \\
  110P/Hartley 3	&2.465&	11.7&	JFC &	4.3	&10.153&\citet{Kokotanekova2017}\\
   14P/Wolf&	2.729	&27.94&	JFC	&4.66	&9.02&\citet{Kokotanekova2017}\\
   10P/Tempel 2&	1.421&	12.03&	JFC &	10.6&	8.93&\citet{Ahearn1989}\\
    &	&	&	&&	&\citet{Jewitt1989} \\
     &	&	&	&&	&\citet{Wisniewski1990} \\
       &	&	&	&&	&\citet{Knight2012} \\
  169P/NEAT	&0.607&	11.3&	JFC	&&	8.369&\\
  137P/Shoemaker-Levy2&	1.933&	4.85&	JFC	&5.8&	7.7&\citet{Kokotanekova2017}\\
  123P/West-Hartley	&2.126&	15.35&	JFC	&&	3.7&\citet{Kokotanekova2017}\\
     C/2003 WT42 (LINEAR)&	5.191&	31.41	&HYP	&&	3.31&\citet{Dermawan2011}\\
\end{tabular}
\label{table:comets}
\end{table*}

\begin{table*}
\centering
\caption{\textbf{Properties of the currently known active asteroids.} This table is reproduced from \citet{Jewitt22_asteroid} with permission from the authors. }
\begin{tabular}{ cccccccc } 
 Object Name & $a$ [au] & $e$ & $i$ [$^\circ$] & $H$ [mag] & $R_{\rm Nuc}$ [m] & $P_{\rm Rot}$ [h] & Reference\\
 \hline
 (1) Ceres	&	2.766	&	0.078	&	10.588	&	3.53	&	469.7	&	9.07 & \citet{Kuppers2014}	\\
	&	&	&		&	&	& & \citet{Park2016}	\\
(493) Griseldis	&	3.116	&	0.176	&	15.179	&	10.97	&	20.78	&	51.94&\citet{Tholen2015}	\\
(596) Scheila	&	2.929	&	0.163	&	14.657	&	8.93	&	79.86	&	15.85 & \citet{Ishiguro2011}	\\
(2201) Oljato	&	2.174	&	0.713	&	2.522	&	15.25	&	0.90	&	$>$26.&\citet{Russell1984}	\\
	&	&	&		&	&	& & \citet{Tedesco2004}	\\
(3200) Phaethon	&	1.271	&	0.890	&	22.257	&	14.32	&	3.13	&	3.60 & \citet{Jewitt2010}	\\
	&	&	&		&	&	& & \citet{Ansdell2014}	\\
(6478) Gault	&	2.305	&	0.193	&	22.812	&	14.81	&	2.8	&	2.49 &\citet{Devogele2021}	\\
(62412) 2000 SY$_{178}$	&	3.159	&	0.079	&	4.738	&	13.74	&	5.19	&	3.33&\citet{Sheppard2015}	\\
(101955) Bennu	&	1.126	&	0.204	&	6.035	&	20.21	&	0.24	&	4.29 &\citet{Lauretta2019}	\\
107P/(4015) Wilson-Harrington	&	2.625	&	0.632	&	2.799	&	16.02	&	3.46	&	7.15 &\citet{Fernandez1997}	\\
	&	&	&		&	&	& & \citet{Licandro2009}	\\
		&	&	&		&	&	& & \citet{Urakawa2011}	\\
133P/(7968) Elst-Pizarro	&	3.165	&	0.157	&	1.389	&	15.49	&	1.9	&	3.47 &\citet{Hsieh2009}	\\
176P/(118401) LINEAR	&	3.194	&	0.193	&	0.235	&	15.10	&	2.0	&	22.23 &\citet{Hsieh2009}	\\
233P/La Sagra (P/2005 JR$_{71}$)	&	3.033	&	0.411	&	11.279	&	16.6	&	1.5	&	— &\citet{Mainzer2010}	\\
238P/Read (P/2005 U1)	&	3.162	&	0.253	&	1.266	&	19.05	&	0.4	&	—& \citet{Hsieh2011b}	\\
259P/Garradd (P/2008 R1)	&	2.727	&	0.342	&	15.899	&	19.71	&	0.30	&	—	&\cite{MacLennan2012}\\
288P/(300163) 2006 VW$_{139}$	&	3.051	&	0.201	&	3.239	&	17.8,18.2	&	0.9,0.6	&	— &\citet{Agarwal2020}	\\
311P/PANSTARRS (P/2013 P5)	&	2.189	&	0.116	&	4.968	&	19.14	&	0.2	&	$>$5.4 &\citet{Jewitt2018}	\\
313P/Gibbs (P/2014 S4)	&	3.154	&	0.242	&	10.967	&	17.1	&	1.0	&	— &\citet{Hsieh2015}	\\
324P/La Sagra (P/2010 R2)	&	3.098	&	0.154	&	21.400	&	18.4	&	0.55	&	— &\citet{Hsieh2015b}	\\
331P/Gibbs (P/2012 F5)	&	3.005	&	0.042	&	9.739	&	17.33	&	1.77	&	3.24&\citet{Drahus2015}	\\
354P/LINEAR (P/2010 A2)	&	2.290	&	0.125	&	5.256	&	—	&	0.06	&	11.36 &\citet{Snodgrass2010}	\\
358P/PANSTARRS (P/2012 T1)	&	3.155	&	0.236	&	11.058	&	19.5	&	0.32	&	— &\citet{Hsieh2018}	\\
426P/PANSTARRS (P/2019 A7)	&	3.188	&	0.161	&	17.773	&	17.1	&	1.2	&	—&	\\
427P/ATLAS (P/2017 S5)	&	3.171	&	0.313	&	11.849	&	18.91	&	0.45	&	1.4	&\citet{Jewitt2019c}\\
432P/PANSTARRS (P/2021 N4)	&	3.045	&	0.244	&	10.067	&	$>$18.2	&	$<$0.7	&	—	\\
433P/(248370) 2005 QN$_{173}$	&	3.067	&	0.226	&	0.067	&	16.32	&	1.6	&	—	&\citet{Hsieh2021MBC}\\
	&	&	&		&	&	& & \citet{Novakovic2022}	\\
P/2013 R3 (Catalina-PANSTARRS)	&	3.033	&	0.273	&	0.899	&	—	&	$\sim$0.2 	&	— &\citet{Jewitt2014}	\\
P/2015 X6 (PANSTARRS)	&	2.755	&	0.170	&	4.558	&	$>$18.2	&	$<$0.7	&	—	&\citet{Moreno2016}\\
P/2016 G1 (PANSTARRS)	&	2.583	&	0.210	&	10.968	&	—	&	$<$0.4	&	—&\citet{Moreno2016b}	\\
		&	&	&		&	&	& & \citet{Hainaut2019}	\\
P/2016 J1-A/B (PANSTARRS)	&	3.172	&	0.228	&	14.330	&	—	&	$<$0.4,$<$0.9	&	— & \citet{Moreno2017}	\\
P/2017 S9 (PANSTARRS)	&	3.156	&	0.305	&	14.138	&	$>$17.8	&	$<$0.8	&	—&\citet{Weryk2017}	\\
P/2018 P3 (PANSTARRS)	&	3.007	&	0.416	&	8.909	&	$>$18.6	&	$<$0.6	&	—	&\citet{Weryk2018}\\
P/2019 A3 (PANSTARRS)	&	3.147	&	0.265	&	15.367	&	$>$19.3	&	$<$0.4	&	—	&\citet{Weryk2019}\\
P/2019 A4 (PANSTARRS)	&	2.614	&	0.090	&	13.319	&	—	&	0.17	&	—	&\citet{Moreno2021}\\
P/2020 O1 (Lemmon-PANSTARRS)	&	2.647	&	0.120	&	5.223	&	$>$17.7	&	$<$0.9	&	—&	\\
P/2021 A5 (PANSTARRS)	&	3.047	&	0.140	&	18.188	&	—	&	0.15	&	—&\citet{Moreno2021}	\\
P/2021 L4 (PANSTARRS)	&	3.165	&	0.119	&	16.963	&	$>$15.8	&	$<$2.2	&	—&	\\
P/2021 R8 (Sheppard)	&	3.019	&	0.294	&	2.203	&	—	&	—	&	—&	\\
\end{tabular}
\label{table:acive_asteroids}
\end{table*}

\end{document}